\documentclass[Crown,sagev,times,super]{sagej}

\usepackage{moreverb,url}

\usepackage{algorithm}
\usepackage{algpseudocode}

\usepackage[colorlinks,bookmarksopen,bookmarksnumbered,citecolor=red,urlcolor=red]{hyperref}

\newcommand\BibTeX{{\rmfamily B\kern-.05em \textsc{i\kern-.025em b}\kern-.08em
T\kern-.1667em\lower.7ex\hbox{E}\kern-.125emX}}

\def\volumeyear{2024}

\begin{document}

\runninghead{Molinari and Thoresen}

\title{A Computationally Efficient Approach to False Discovery Rate Control and Power Maximisation via Randomisation and Mirror Statistic}

\author{Marco Molinari\affilnum{1} and Magne Thoresen\affilnum{1}}

\affiliation{\affilnum{1}University of Oslo, Oslo Center for Biostatistics}
\corrauth{Marco Molinari, 
Oslo Center for Biostatistics,
University of Oslo,
P.O.Box 1122 Blindern, 0317 Oslo,
Norway.}
\email{marco.molinari@medisin.uio.no}

\begin{abstract}
Simultaneously performing variable selection and inference in high-dimensional regression models is an open challenge in statistics and machine learning. The increasing availability of vast amounts of variables requires the adoption of specific statistical procedures to accurately select the most important predictors in a high-dimensional space, while controlling the False Discovery Rate (FDR) associated with the variable selection procedure. In this paper we propose the joint adoption of the Mirror Statistic approach to FDR control, coupled with outcome randomisation to maximise the statistical power of the variable selection procedure, measured through the True Positive Rate (TPR). Through extensive simulations we show how our proposed strategy allows to combine the benefits of the two techniques. The Mirror Statistic is a flexible method to control FDR, which only requires mild model assumptions, but requires two sets of independent regression coefficient estimates, usually obtained after splitting the original dataset. Outcome randomisation is an alternative to Data Splitting, that allows to generate two independent outcomes, which can then be used to estimate the coefficients that go into the construction of the Mirror Statistic. The combination of these two approaches provides increased testing power in a number of scenarios, such as highly correlated covariates and high percentages of active variables. Moreover, it is scalable to very high-dimensional problems, since the algorithm has a low memory footprint and only requires a single run on the full dataset, as opposed to iterative alternatives such as Multiple Data Splitting.
\end{abstract}


\maketitle

\section{Introduction}\label{sec:Intro}
Advances in data collection capabilities have allowed researchers to get access to thousands of features on multiple subjects in relatively short times. Examples are next-generation sequencing technologies that allow fast DNA and RNA sequencing and Nuclear Magnetic Resonance Spectroscopy used to identify metabolites. \cite{Borah2024} But also wearable personal devices that can measure several anthropometric variables and clinical outcomes of interests, such as blood pressure and blood glucose levels, continuously over time, thus producing a vast collection of measurements. \cite{Berry2020} Moreover, it is now common to integrate these multiple input sources into a single study, and repeating these measurements over multiple time points, in a longitudinal fashion, in order to capture potentially relevant trends in the outcomes of interest, given some specific interventions or treatments.\\
The combination of all these factors generates an incredibly vast and complex set of features that need to be jointly analyzed. \cite{Konietschke2020, Rahnenfuehrer2023} In many such studies, feature selection becomes an important task, a typical example being biomarker discovery. This can be a daunting task, for at least two reasons: $1)$ the available sample size, e.g. the number of patients recruited in a cohort study, is very often much lower than the number of available features (high-dimensional problem). Therefore, standard statistical or machine learning models struggle to extract the underlying fundamental associations, due to model limitations. $2)$ even when a model can deal with a sample size lower than the number of features, the high probability of selecting false positive effects, poses a serious threat to the validity of the analysis.\\
We address the first problem of variable selection by using all available features during the training process, building a statistical model that can automatically select the most relevant variables. This allows to retain the full interpretability about the effects of the covariates on the outcome of interest, in contrast with dimensionality reduction techniques such as Principal Component Analysis.\citep{Ayesha2020} In high-dimensional regression (or classification) problems, popular choices are the LASSO, ElasticNet, LARS and SCAD, which provide efficient algorithms that scale well to a large number of features.\cite{Tibshirani1996, Zou2005, Efron2004, Fan2011} Alternatives also exist in the Bayesian framework, where variable selection is performed through an appropriate choice of the prior distributions.\cite{OHara2009} Popular choices are the two-group Spike and Slab prior distribution, which explicitly provide posterior probabilities of inclusion for each variable; and the one-group prior, such as the Laplace prior distribution, which shrinks coefficients toward zero, acting as a regularisation in a similar fashion to the $L_1$ penalty in the LASSO.\\
In addition to selecting only a subset of variables, in many applications to real data it is also essential to be able to make proper inference on the selected subset of features. This is essential to have a reliable estimate of the regression coefficients confidence intervals and to be able to control some form of error rate, such as the False Discovery Rate (FDR).\cite{Benjamini1995} However, simply proceeding with inference, after a data-dependent variable selection step, does not allow to perform valid \textit{post-selection} inference, as explained in Berk et al.\cite{Berk2013} This is because data-driven variable selection procedures, such as the ones mentioned above, generate a model that is not deterministic and the straightforward application of classic approaches to inference, like Ordinary Least Squares, do not account for this additional randomness. Therefore, the estimated coefficients, and the corresponding confidence intervals will be biased, leading to a potential increase in erroneous classifications.\\
A number of methods exist to control error rates in multiple testing. Meinshausen and Buhlmann\cite{Meinshausen2010} propose \textit{stability selection}, as a method to control the number of False Discoveries in a LASSO regression setting. Lee et al.\cite{Lee2016} takes a different approach, providing an analytical solution for the linear regression problem when using a LASSO penalty, formally accounting for the variable selection process at the inferential step, thus obtaining correct confidence intervals for the selected coefficients. Rugamer, Baumann and Greven\cite{Ruegamer2022} provide a partial extension to the case of additive and linear mixed models, using Monte Carlo approximations, however the proposed solution is computationally intensive. Overall, this formal treatment is limited to simple cases, such as linear regression, and extensions are difficult.\\
Barber and Candes,\cite{Barber2015} in their pioneering work, propose the \textit{knockoff} filters procedure as a direct way to control the False Discovery Rate (FDR). The knockoff method works by augmenting the space of covariates, adding a perturbed version of each feature to the design matrix and performing variable selection on the new feature space. The authors provide an upper-bound estimate of the FDR and a new knockoff test statistic through which it is possible to control the FDR at any specific level. This approach, however, has some limitations, i.e. the knowledge of the joint distribution of the covariates is required to construct the knockoff filters and the procedure is limited to the case $n > p$. Candes et al.\cite{Candes2018} provide an extension to the high-dimensional scenarios, but still requires knowledge of the joint distribution of the covariates, which is not known in most real data applications.\\
Despite the limitations of the knockoff, the idea of feature perturbation has been successfully adopted in other works as a way to control FDR. Xing, Zhao and Liu\cite{Xing2021} develop the Gaussian Mirrors procedure which allows to control FDR without requiring any distributional assumption on the covariates. However, this approach is inefficient because it only evaluates one variable at a time. Using a similar approach, Dai et al.\cite{Dai2022} develop the Mirror Statistic, by substituting the feature perturbation step with a two-steps procedure based on data splitting (DS). The first step being variable selection on a subset of the data and the second step being statistical inference on a second subset of data, independent of the first. Since the two sets are independent by construction, the conditional distribution of the inference set given the output of the selection step is the same as the unconditional one, so classical procedures can be used to provide valid inference for the selected parameters.\cite{Rasines2022} One drawback of this approach is the loss of power caused by the reduced sample size available for the inference step (and similarly for the variable selection step). To mitigate this problem the authors propose a variation called Multiple Data Splitting (MDS), more akin to stability selection, showing increased power in simulations.\cite{Dai2022} The authors show that this approach outperforms the knockoff filter in many situations. Nevertheless, MDS is computationally much more costly than DS, since the same procedure has to be repeated multiple times (at least $50$ according to the authors).\\

In this paper we propose to use the Mirror Statistic, but, instead of creating two independent sub-samples using data splitting, we borrow the idea of outcome randomisation from Rasines and Young\cite{Rasines2022}, where the authors propose a simple mechanism to create two independent sets of data by adding some random noise to the outcome, splitting more efficiently the original information available into two independent new pseudo-outcomes. The result is an increase in statistical power and a more computationally efficient algorithm. We provide a performance comparison via numerical experiments, replicating the results of Dai et al.\cite{Dai2022} and extending the simulation scenarios to more challenging settings.\\
Throughout the article we use the following notation: RandMS to indicate our proposed model with outcome Randomisation plus Mirror Statistic, DS for single Data Splitting with Mirror Statistic, MDS for Multiple Data Splitting with Mirror Statistic. $S_0$ indicates the set of features with a true null coefficient and $S_1$ the set of features with a true non-null coefficient (active variables). $\text{X}$ denotes the whole matrix of covariates, $\text{X}_j$ denotes column $j$ and $\boldsymbol{y}$ denotes the vector outcome. We make use of the terms variables, covariates and features interchangeably.\\

The remainder of the article is organized as follows. In Section \ref{sec:Methods} we review the methodology underlying FDR control via the Mirror Statistic and Data Splitting. We then introduce Randomisation and provide the algorithm for our proposed strategy. In Section \ref{sec:Simulations} we show in detail the results of our simulations and the computational performance of the algorithm. In Section \ref{sec:Application} we apply the proposed method to the selection of genes in a high-dimensional real-world study. Finally, in Section \ref{sec:Discussion} we summarise our contributions, limitations and potential extensions of the method.

\section{Methods}\label{sec:Methods}

\subsection{False Discovery Rate control}
The False Discovery Rate has been introduced as a less conservative approach to False Positive error control compared to the Family-Wise Error Rate, which can be too restrictive when testing a large number of hypothesis.\cite{Benjamini1995} FDR is defined as the expectation of the False Discovery Proportion:
\begin{align}\label{eq:fdr}
    \text{FDR} = \mathbf{E}\left[ \text{FDP} \right] = \mathbf{E}\left[ \dfrac{\#\left\{j:  j \in S_0, j \in \hat{S} \right\}}{\#\left\{ j: j \in \hat{S} \right\} \vee 1 } \right]
\end{align}
where the expectation is taken with respect to the stochastic model selection procedure and the randomness in the data. $\hat{S}$ represents the set of all selected features.\\
Benjamini and Hochberg\cite{Benjamini1995} provide a correction method for the p-values so that a specific FDR level $\alpha$ can be achieved, under the assumption of independence of the p-values. Although some extensions exist that allow for some form of dependence,\cite{Benjamini2001} for many of the aforementioned algorithms in Section \ref{sec:Intro}, p-values are not available at all (e.g. LASSO). Hence, the necessity of using methods that can achieve FDR control without explicitly calculating p-values, such as the Mirror Statistic (Equation \ref{eq:mirror_stat}).\\
For completeness we also specify the formula for the True Positive Rate (TPR), defined as
\begin{align*}
    \text{TPR} = \dfrac{\#\left\{j:  j \in S_1, j \in \hat{S} \right\}}{\#\left\{ j: j \in S_1 \right\} }
\end{align*}
The TPR is used to measure the power of the method, i.e. the ability to select the true active variables.

\subsection{Data Splitting and Randomisation}
The practice of splitting a given dataset into multiple smaller independent slices is common and the standard in most machine learning applications, \cite{Bishop2006} where, generally, a training, validation and testing set are generated from the original sample. While in machine learning, splitting is done to validate the prediction accuracy of a model, in classic statistical inference the same idea can be used to create valid inferential procedures. Dai et al.\cite{Dai2022} use this strategy in order to control FDR via the test statistic Mirror Statistic, defined for variable $\text{X}_j$ as:
\begin{align}\label{eq:mirror_stat}
    M_j = \text{sign}\left(\ \hat{\beta}^{(1)}_j \hat{\beta}^{(2)}_j \right)
        f \left( |\hat{\beta}^{(1)}_j| , |\hat{\beta}^{(2)}_j| \right)
\end{align}
where $f(|x|,|y|)$ is a non-negative, exchangeable and monotonically increasing function and $\hat{\beta}^{(1)}_j$ and $\hat{\beta}^{(2)}_j$ are two distinct estimates of the regression coefficient obtained on two independent subsets of the sample. The logic behind Equation \ref{eq:mirror_stat} is that for features that are relevant, the corresponding $M_j$ will get a positive relatively large value, because the two independent estimates, $\hat{\beta}^{(1)}_j$ and $\hat{\beta}^{(2)}_j$, will have concordant signs. Conversely, if the estimated coefficients have discordant signs, $M_j$ will always get a negative value and if the estimates are relatively small, meaning that probably the feature is not relevant, then $M_j$ will be small as well.\\
Under the assumption that, for a feature $j \in S_0$, the sampling distribution of at least one of the two coefficients is symmetric around zero, then also $M_j$ will be symmetric around zero. This property, plus the Mirror Statistic construction, provides an upper bound on the number of false positives:
\begin{align}\label{eq:fp_approximation}
\begin{split}
    &\#\left\{j \in S_0: M_j > t \right\} \\
    \approx \ &\#\left\{j \in S_0: M_j < -t \right\} \\
    \leq \ &\#\left\{j: M_j < -t \right\}, \forall t > 0
\end{split}
\end{align}
which can be directly used to approximate the False Discovery Proportion in Equation \ref{eq:fdr}.\\
As we can see from the definition, to use the Mirror Statistic estimator we need two independent sets of observations. Although data splitting is universally valid and straightforward to use, it comes at the cost of a much reduced sample size, which can have detrimental effects on the power of the statistical test and on the stability of the variable selection. To counteract this downside Dai et al.\cite{Dai2022} propose MDS as a way to increase the power of the test statistic. MDS amounts to repeating multiple times the whole procedure of variable selection with simple DS and then aggregating the results. In simulations MDS seems to provide higher power; however, this improvement comes at the cost of a much higher computational burden and an additional uncertainty due to the choice of the aggregation strategy.\\
This is where Randomisation offers an alternative way of distributing the available sample information and helps avoiding the randomness of the simple data splitting, creating two pseudo-independent sets, by perturbing the outcome with some random noise $W$.\cite{Rasines2022} The general idea is to use a randomisation scheme through which we only allow ourselves to observe the outcome through $U \equiv u(Y, W)$ at the variable selection step, while at the inference step we only observe $V \equiv v(Y, W)$, with $V$ constructed to be independent of $U$.\\
Here we consider the case where the outcome of interest has a Normal distribution and whose mean is a function of some features $\text{X}$.
\begin{align}\label{eq:linear_model}
    \boldsymbol{y} \sim \text{N}_n(\boldsymbol{\mu} = g(\text{X}), \sigma^2 \text{I}_n)
\end{align}
where $\text{I}_n$ is the $n$-dimensional identity matrix.\\
Given $\sigma^2$, or an estimate $\hat{\sigma}^2$, we can generate an $n$-dimensional vector of random Normal noise as $\boldsymbol{w} \sim \text{N}_n(\boldsymbol{0}, \sigma^2 \gamma \text{I}_n)$, where the scalar $\gamma > 0$ allows to distribute information for variable selection and inference, respectively. Rasines and Young\cite{Rasines2022} show how using $\gamma = 1$ is equivalent to splitting the data in two halves of equal size, which is the most natural choice in absence of any additional information. Any other choice would place more information on either the LASSO or the OLS.\\
Then, building $U = Y + W$, we have that $\boldsymbol{u} \sim \text{N}_n(\boldsymbol{\mu}, \sigma^2(1 + \gamma)\text{I}_n)$ and $V = Y - \gamma^{-1}W$, with $\boldsymbol{v} \sim \text{N}_n(\boldsymbol{\mu}, \sigma^2(1 + \gamma^{-1})\text{I}_n)$.\\
Randomisation can be interpreted as averaging information over all possible data splits of the same size. Rasines and Young\cite{Rasines2022} show that for a Normally distributed outcome $\boldsymbol{y}$, randomisation guarantees a power that is always at least as high as data-splitting.\\
The complete inferential procedure is detailed in the following Algorithm:
\begin{algorithm}
\caption{Randomisation + Mirror Statistic}\label{alg:randms}
\begin{algorithmic}[1]
    \Require $\gamma$, $q$ (FDR target level), ${\sigma}^2$
        \State $\boldsymbol{w} \overset{IID}{\sim} \text{N}_n(\boldsymbol{0}, \sigma^2 \gamma \text{I}_n)$
        \State $\boldsymbol{u} = \boldsymbol{y} + \boldsymbol{w}$
        \State $\boldsymbol{v} = \boldsymbol{y} - \gamma^{-1}\boldsymbol{w}$\\
        % Mirror Stat
        Estimate regression coefficients $\boldsymbol{\hat{\beta}^{(u)}}$ and $\boldsymbol{\hat{\beta}^{(v)}}$ from $\boldsymbol{u}$ and $\boldsymbol{v}$ respectively.\\
        Calculate MS $M_j$, $\forall j$, using Equation \ref{eq:mirror_stat}.\\
        Select the covariates $\text{X}_j$ for which ${M_j > \tau_q}$, where the threshold $\tau_q$ is
        $$
            \tau_q = \text{min}\left\{ t>0 : \text{FDP}(t) = \dfrac{\#\left\{j:  M_j < -t \right\}}{\#\left\{ j: M_j > t \right\} \vee 1 } \leq q \right\}
        $$
    \end{algorithmic}
\end{algorithm}

If the residual variance ${\sigma}^2$ is unknown, it can be estimated using the residual sum of squares obtained from the LASSO model fitted with penalisation parameter tuned by $k$-fold cross-validation,\cite{Rasines2022}, that is:
\begin{align*}
    \hat{\sigma}^2 = \sum_{i=1}^{\tilde{n}} \dfrac{\left( y_i - \hat{y}_i \right)^2}{\tilde{n} - df - 1}
\end{align*}
where $\tilde{n}$ is the sample size, $df$ is the number of non-zero coefficients and $\hat{y}$ is the predicted outcome.\\
This is the strategy that we adopt in the numerical experiments in Section \ref{sec:Simulations}, as well as in the real-world data application in Section \ref{sec:Application}, where in both cases we use $k=10$ folds.\\
We further explore the implications of an incorrectly estimated variance ${\sigma}^2$ through additional simulations where we control the value of ${\sigma}^2$ used in the model. In the Appendix, we report the results of this experiment. When the variance is underestimated the FDR is not properly controlled, while when the variance is overestimated the algorithm becomes more conservative in terms of FDR control.

\subsection{Assumptions}
The theory underlying the development of the Mirror Statistic is based on the independence of the two sets of coefficient estimates and the following \textit{symmetry} and \textit{weak dependence} assumptions:\cite{Dai2022}
\begin{enumerate}
    \item \textit{Symmetry}: For each feature $j \in S_0$, the sampling distribution of at least one of the two coefficients, $\hat{\beta}^{(1)}_j$ and $\hat{\beta}^{(2)}_j$, is symmetric around zero
    \item \textit{Weak dependence}: The mirror statistics $M_j$ are continuous random variables, and there exist a constant $c > 0$ and $\alpha \in (0, 2)$ such that
    $$
    \text{Var}\left( \#\left\{M_j > t \mid j \in S_0\right\} \right) \leq c p_0^{\alpha}, \forall t > 0
    $$
    where $p_0$ is the number of null features. This assumption translates into a restriction on the correlation among the null features

\end{enumerate}
A requirement for the \textit{symmetry} condition is that all variables with a non-null coefficient are selected in the variable selection step. This is generally referred to as the \textit{sure screening} property. In our simulations and real-world application we use the combination of LASSO and OLS, respectively to perform variable selection and inference, in linear models. For the LASSO, the \textit{sure screening} property depends on the \textit{Signal strength condition}, defined as:
$$
\textit{Signal strength condition}: \underset{j \in S_1}{\text{min}} |\beta_j| \gg \sqrt{p_1 \log p / n}
$$
In order to better understand the impact of violating the assumption of symmetry we run a controlled simulation where we monitor the individual steps of the randomization plus Mirror Statistic algorithm. Results are detailed in the Appendix.

\section{Simulations}\label{sec:Simulations}
To prove the effectiveness of our proposed approach we perform several simulations in multiple scenarios, comparing our method against DS and MDS.\\
We first repeat the simulations done in Dai et al.\cite{Dai2022}, to check whether we can achieve a similar performance to what was reported in the original paper. We then perform additional simulations under different scenarios, with the purpose of finding the limit of the approach, i.e. when the performance deteriorates too much, and to check in which situations our method can perform better than the alternatives.\\
The common strategy to all simulations is to perform variable selection with LASSO and inference with a standard linear model. For all simulations we set $f(a, b) = a + b$ in Equation \ref{eq:mirror_stat}.\cite{Dai2022}\\
The metrics that we track are the False Discovery Rate (FDR) and the True Positive Rate (TPR, or Power). An ideal model selection will have a TPR close to $1$ for every level of FDR we wish to control for.

\subsection{Replication of the simulations from the original paper with DS and MDS}\label{paper_simu_replica}

The data for this scenario is generated using the following parameters:
\begin{itemize}
    \item sample size, $n = 800$
    \item number of covariates, $p = 2000$
    \item number of non-zero coefficients, $p_1 = 50$, equivalent to $2.5\%$ of $p$
    \item covariates correlation, $\rho = \left\{0, 0.2, 0.4, 0.5, 0.6, 0.8 \right\}$
    \item regression coefficients signal strength, $\delta = \left\{3, 4, 5, 6, 7 \right\}$
    \item Error variance $\sigma^2 = 1$
    \item Randomization variance factor $\gamma = 1$
    \item Predictor in Equation \ref{eq:linear_model} set to be $g(\textbf{X}) = \textbf{X} \boldsymbol{\beta}$
\end{itemize}
Throughout the paper, the correlation coefficient $\rho$ represents the highest value from which the correlation matrix is built, following the structure defined in Equation \ref{eq:sigma_toeplitz}, unless otherwise specified.\\
The covariates are generated as independent random draws from a multivariate Normal distribution, $\boldsymbol{x}_i \sim \text{N}_p(\textbf{0}, \Sigma)$, where the covariance $\Sigma$ is constructed as a diagonal Toeplitz matrix, with each block defined as:
\begin{align}\label{eq:sigma_toeplitz}
    \begin{bmatrix}
        1 & \dfrac{(p^{\prime} - 2)\rho}{(p^{\prime} - 1)} & \dfrac{(p^{\prime} - 3)\rho}{(p^{\prime} - 1)} & \dots & \dfrac{\rho}{(p^{\prime} - 1)} & 0\\
        \dfrac{(p^{\prime} - 2)\rho}{(p^{\prime} - 1)} & 1 & \dfrac{(p^{\prime} - 2)\rho}{(p^{\prime} - 1)} & \dots & \dfrac{2\rho}{(p^{\prime} - 1)} & \dfrac{\rho}{(p^{\prime} - 1)}\\
        \vdots & & & \dots & & \vdots\\
        0 & \dfrac{\rho}{(p^{\prime} - 1)} & \dfrac{2\rho}{(p^{\prime} - 1)} & \dots & \dfrac{(p^{\prime} - 2)\rho}{(p^{\prime} - 1)} & 1
    \end{bmatrix}
\end{align}
where $p^{\prime}$ is the dimension of the block.\\
The regression coefficients are randomly generated from a Normal distribution with mean zero and standard deviation $\delta \sqrt{\log(p)/n}$.\\

\begin{figure}
    \centering
    \includegraphics[width=0.8\linewidth]{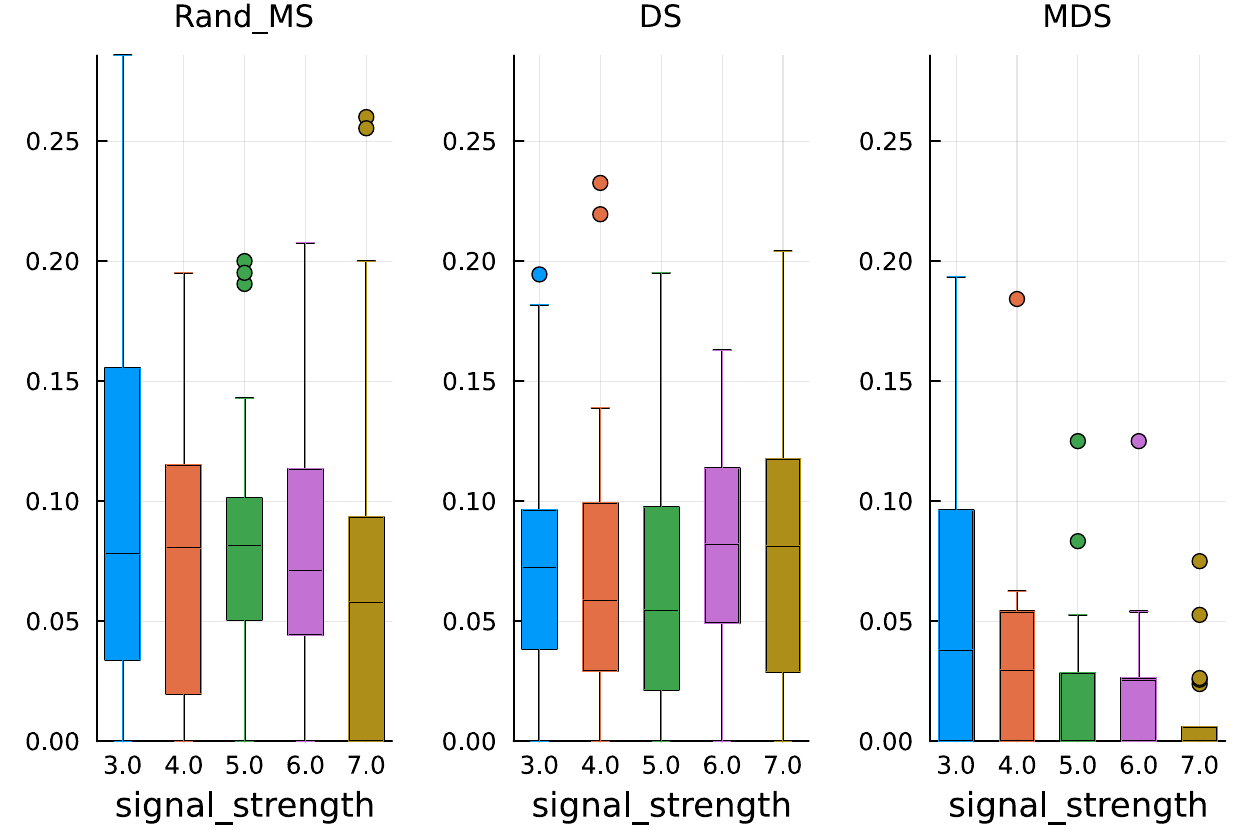}

    \vspace{0.5cm}
    
    \includegraphics[width=0.8\linewidth]{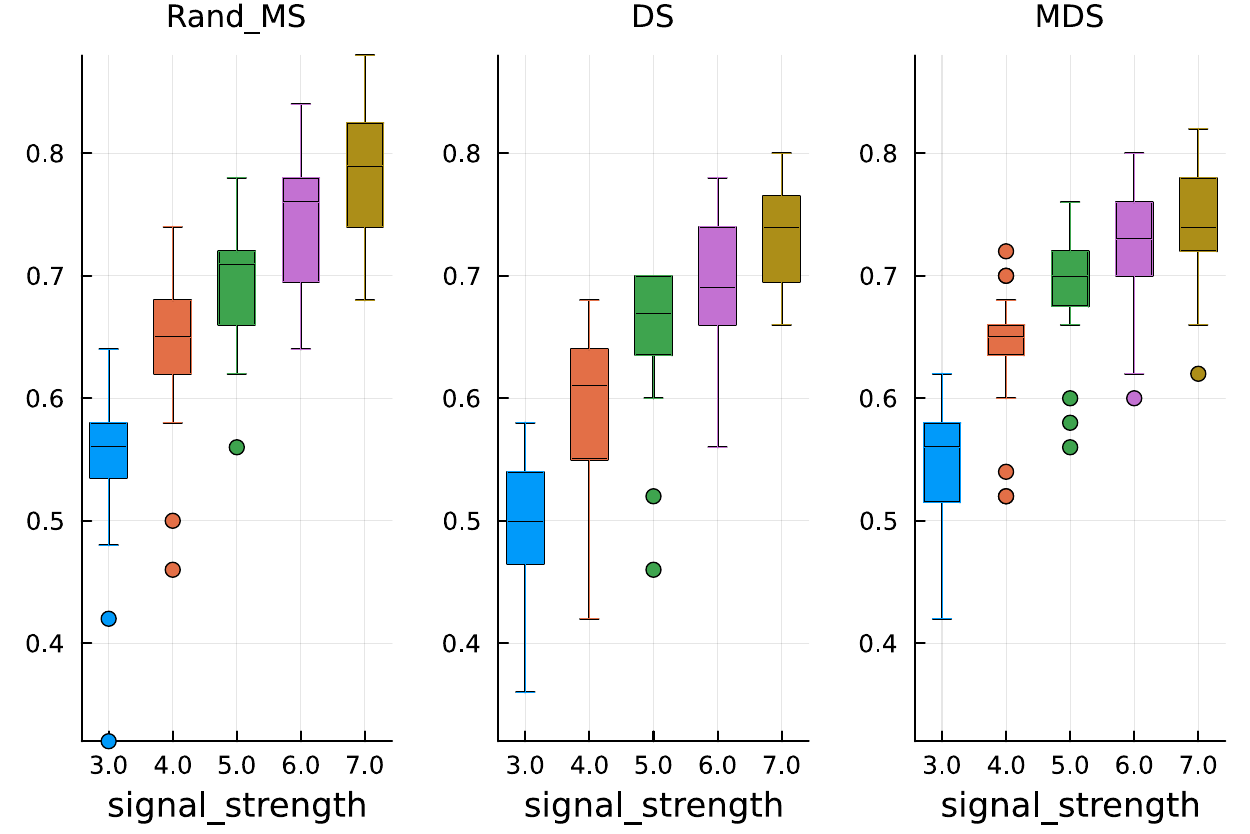}
    \caption{\textcolor{blue}{Paper replication study} - $\rho=0.5$\\
        Top: False Discovery Rate\\
        Bottom: True Positive Rate}
    \label{fig:rep_paper_fdr_rho_5}
\end{figure}

\begin{figure}
    \centering
    \includegraphics[width=0.8\linewidth]{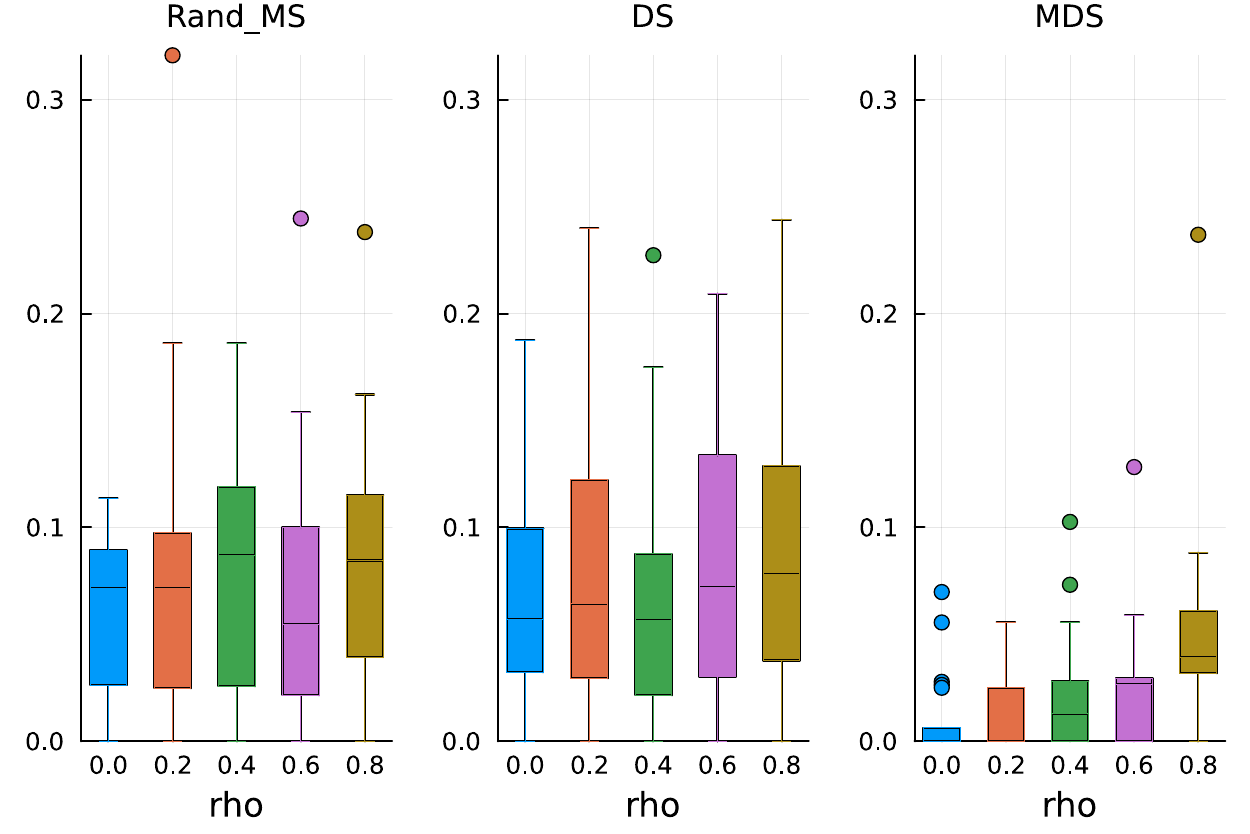}

    \vspace{0.5cm}

    \includegraphics[width=0.8\linewidth]{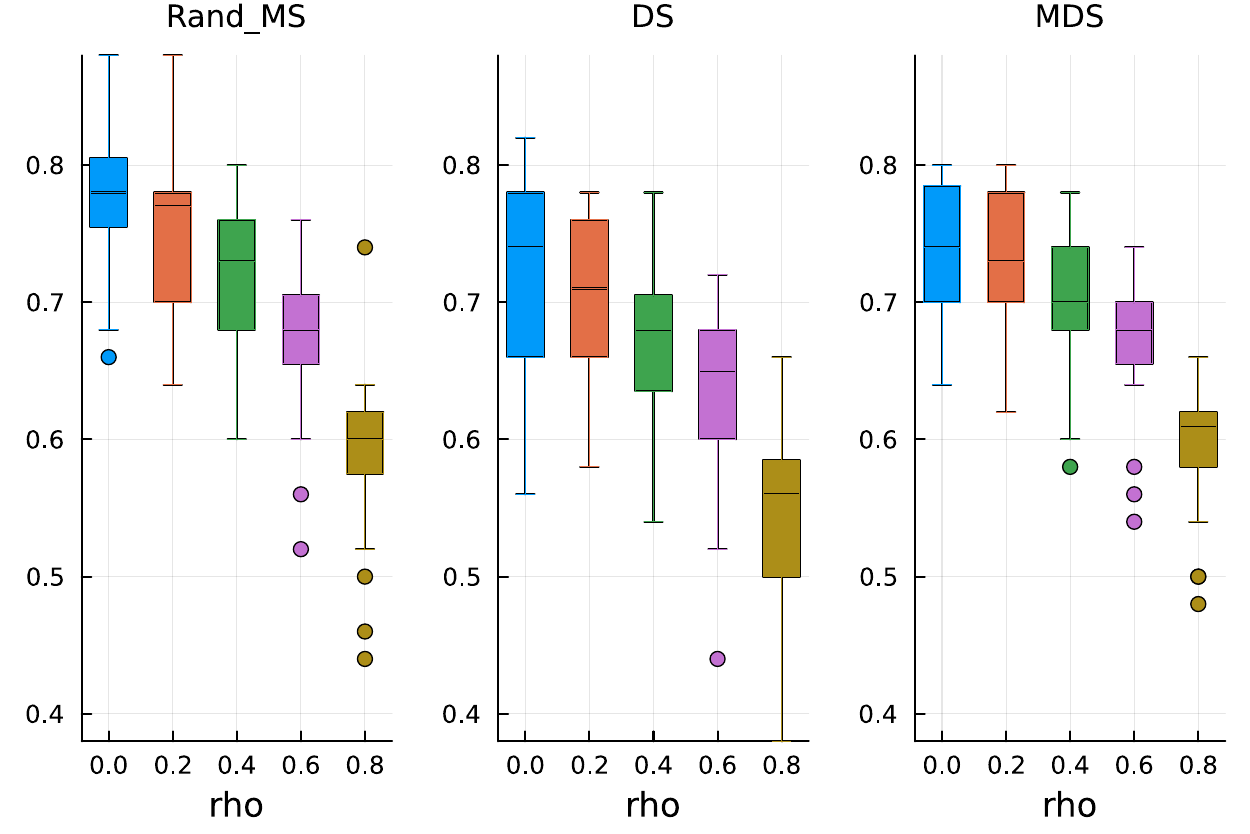}

    \caption{\textcolor{blue}{Paper replication study} - signal strength $\delta=5$\\
        Top: False Discovery Rate\\
        Bottom: True Positive Rate}
    \label{fig:rep_paper_fdr_beta_5}
\end{figure}

Replicating the strategy used by Dai et al.\cite{Dai2022} we first compare the performance of the three algorithms fixing the signal strength to $\delta = 5$, varying the degree of correlation $\rho$, and then fixing the correlation to $\rho = 0.5$ and varying the signal strength. MDS is run using $50$ replications of DS, as recommended by the authors.\\
$50$ independent replications are run for each combination. Here we show the boxplot of the results.\\
In Figures \ref{fig:rep_paper_fdr_rho_5} and \ref{fig:rep_paper_fdr_beta_5}, top plots, we can observe some common patterns: $a)$ all three algorithms achieve, on average, FDR control at the nominal level of $0.1$; $b)$ MDS is always more conservative, achieving on average a lower FDR.\\
In the bottom plots, we show the TPR estimates: $a)$ MDS does not seem to be significantly better than the simple DS; $b)$ RandMS achieves a performance comparable to DS and MDS, with often higher (better) median values.

\subsection{New simulations}
We now concentrate on testing our method on new scenarios that were not covered in the original paper. We explore the performance in contexts with a higher correlation, near-ill conditioned covariance matrices, higher proportions of non-zero regression coefficients, non-block diagonal covariance matrices and regression coefficients sampled from a fixed known pool of values.

\subsubsection{$\beta$ from fixed pool}\label{sebsec:fixed_beta}
\begin{figure}
    \centering
    \includegraphics[width=1.\linewidth]{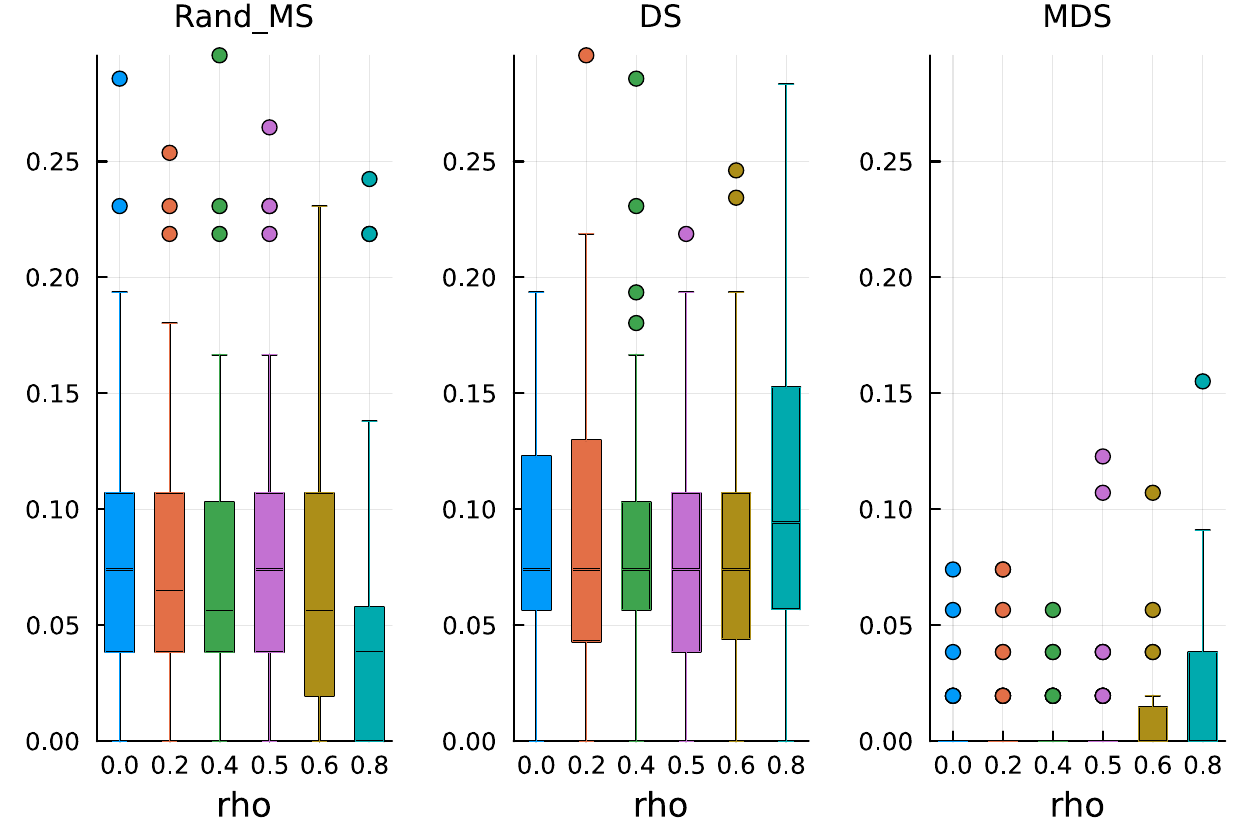}

    \vspace{0.5cm}
    
    \includegraphics[width=1.\linewidth]{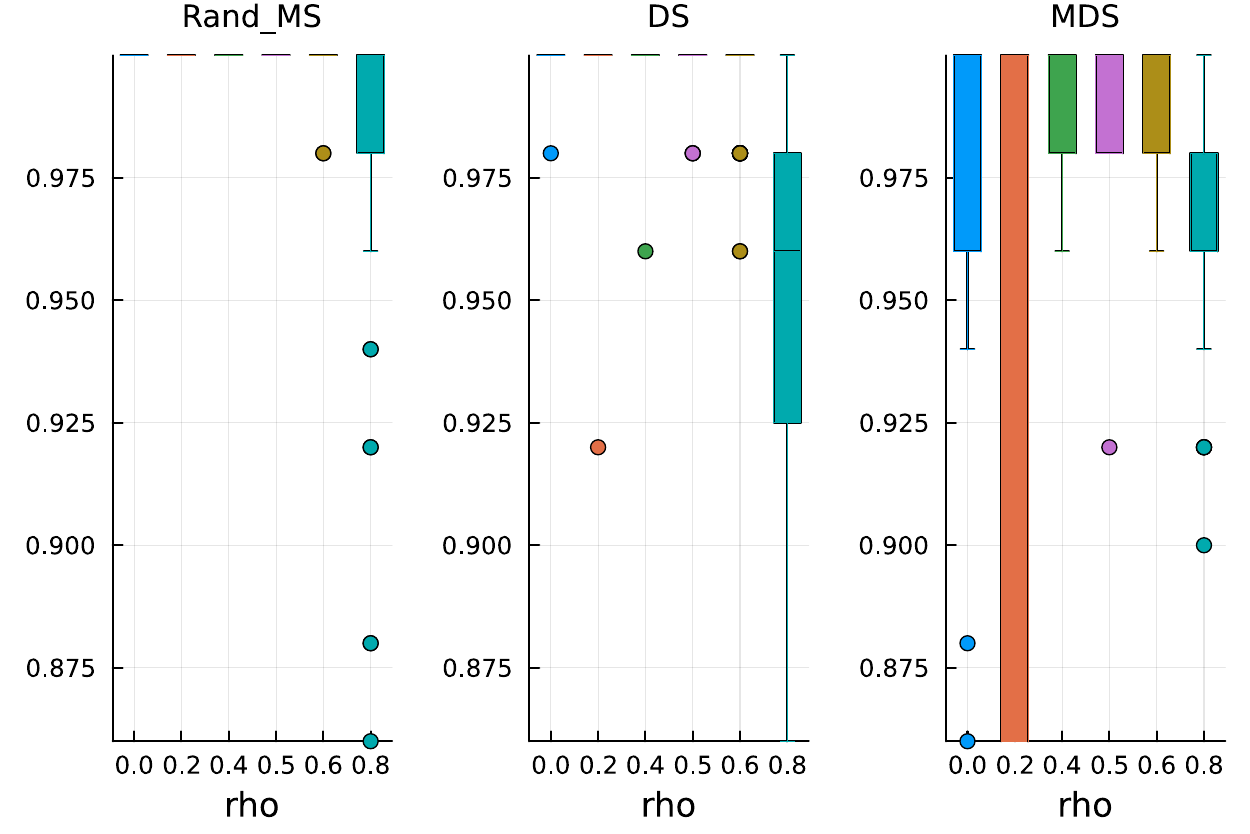}
    
    \caption{\textcolor{blue}{$\beta$ from fixed pool}\\
        Top: False Discovery Rate\\
        Bottom: True Positive Rate}
    \label{fig:beta_fixed_p2000}
\end{figure}

The first additional scenario that we test is the situation where the regression coefficients $\boldsymbol{\beta}$ are not drawn from a distribution, as done in \ref{paper_simu_replica}, but rather randomly sampled from a fixed known pool of values. For this simulation we sample $p$ values from the set $\left\{ -1, -0.8, -0.5, 0.5, 0.8, 1 \right\}$.\\
Selecting the coefficients from a known set of values allows to better control and understand the variable selection capabilities of the algorithms, since the random draws from a Normal distribution will naturally be concentrated around the zero mean, making it more difficult to really understand which coefficients can actually be considered non-zero. This set of regression coefficients is used in all of the following simulations.\\ 
Keeping all the other simulation settings as before, we run the same simulations. From Figure \ref{fig:beta_fixed_p2000} (top) we see that the median values and variability of FDR are very similar to the ones obtained before, validating the ability of the Mirror Statistic to control the FDR. On the other hand, the results for the TPR in Figure \ref{fig:beta_fixed_p2000} (bottom) suggest that variable selection with fixed coefficients is somewhat easier, as expected, which is reflected in higher median TPR values, concentrated near $1$.\\
All three methods have a comparable performance, both in terms of FDR control and TPR, with the only exception of MDS, which has, on average, a lower TPR than DS and RandMS. This behaviour could be explained by the fact that MDS has a very conservative control over FDR, thus also ending up with a lower TPR.

\subsubsection{Higher percentages of non-zero coefficients}
\begin{figure}
    \centering
    \includegraphics[width=1\linewidth]{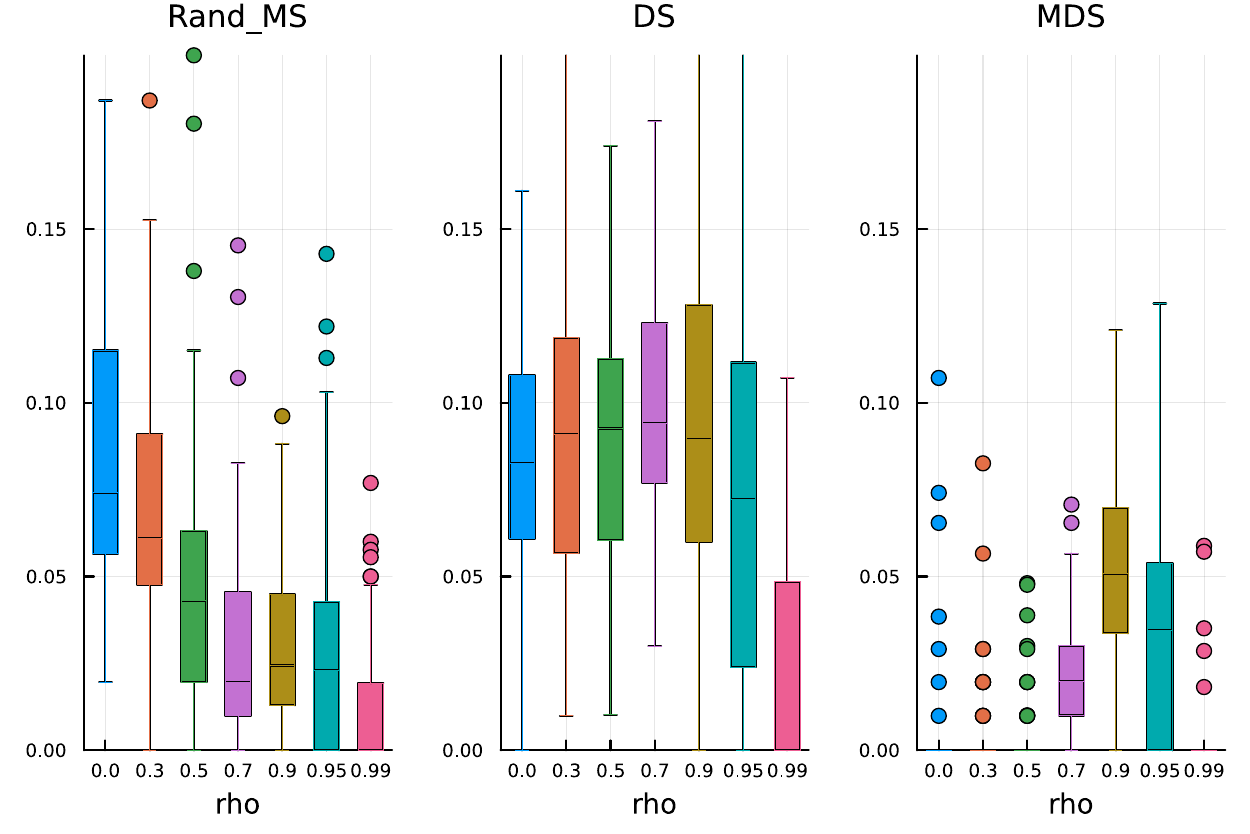}

    \vspace{0.5cm}
    
    \includegraphics[width=1\linewidth]{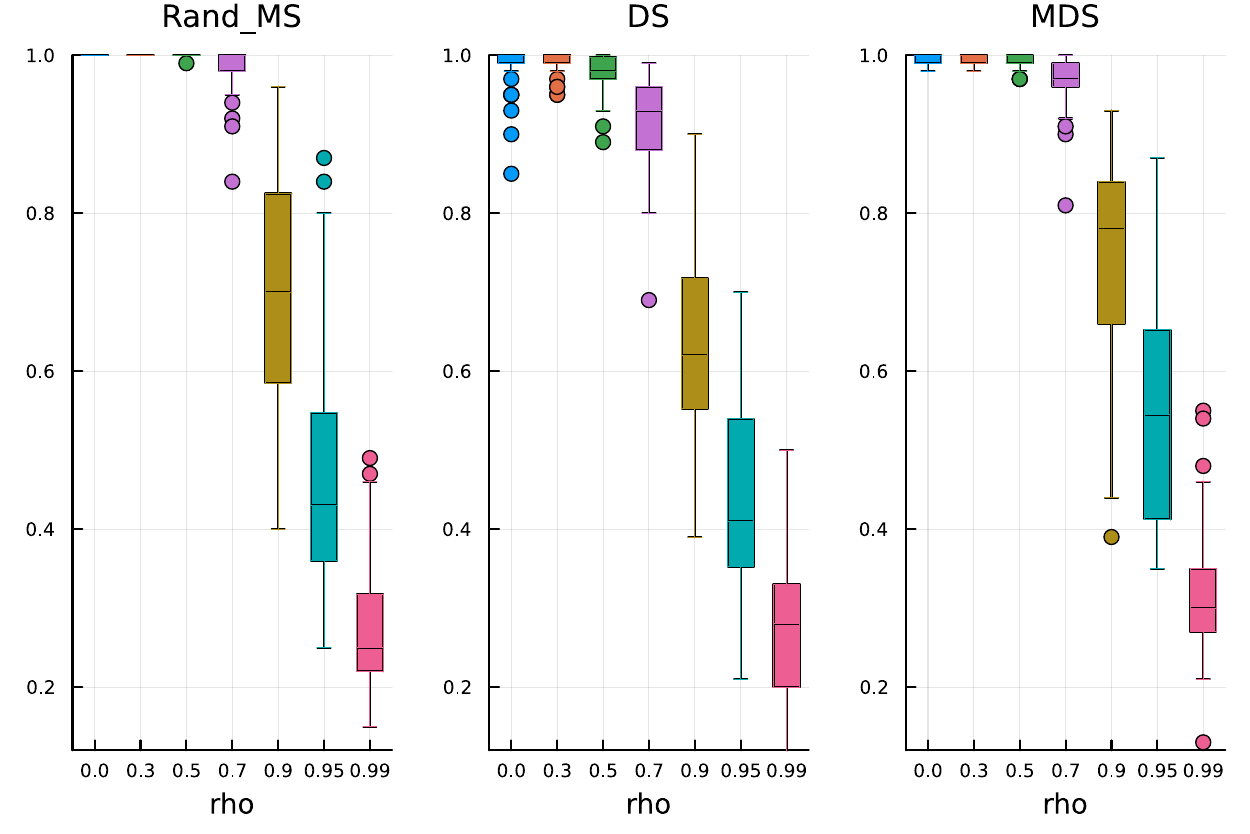}
    \caption{\textcolor{blue}{Active coefficients $10\%$}\\
        Top: False Discovery Rate\\
        Bottom: True Positive Rate}
    \label{fig:p10_lasso}
\end{figure}

A natural question arises regarding whether DS, MDS and RandMS, can cope with a higher percentage of active variables, potentially highly correlated. To this end, we increase the complexity of the simulated data by increasing the percentage of active variables and allowing very high degrees of correlations across the covariates.\\
We start by increasing the percentage of active variables to $10\%$. From Figure \ref{fig:p10_lasso} (top) we see that the FDR is still under control and MDS is conservative, as before. In Figure \ref{fig:p10_lasso} (bottom) we can observe that RandMS is achieving a TPR always at least comparable to DS and MDS. Common to all three methods is the sharp decrease in TPR for higher correlation structures.

\begin{figure}
    \centering
    \includegraphics[width=1\linewidth]{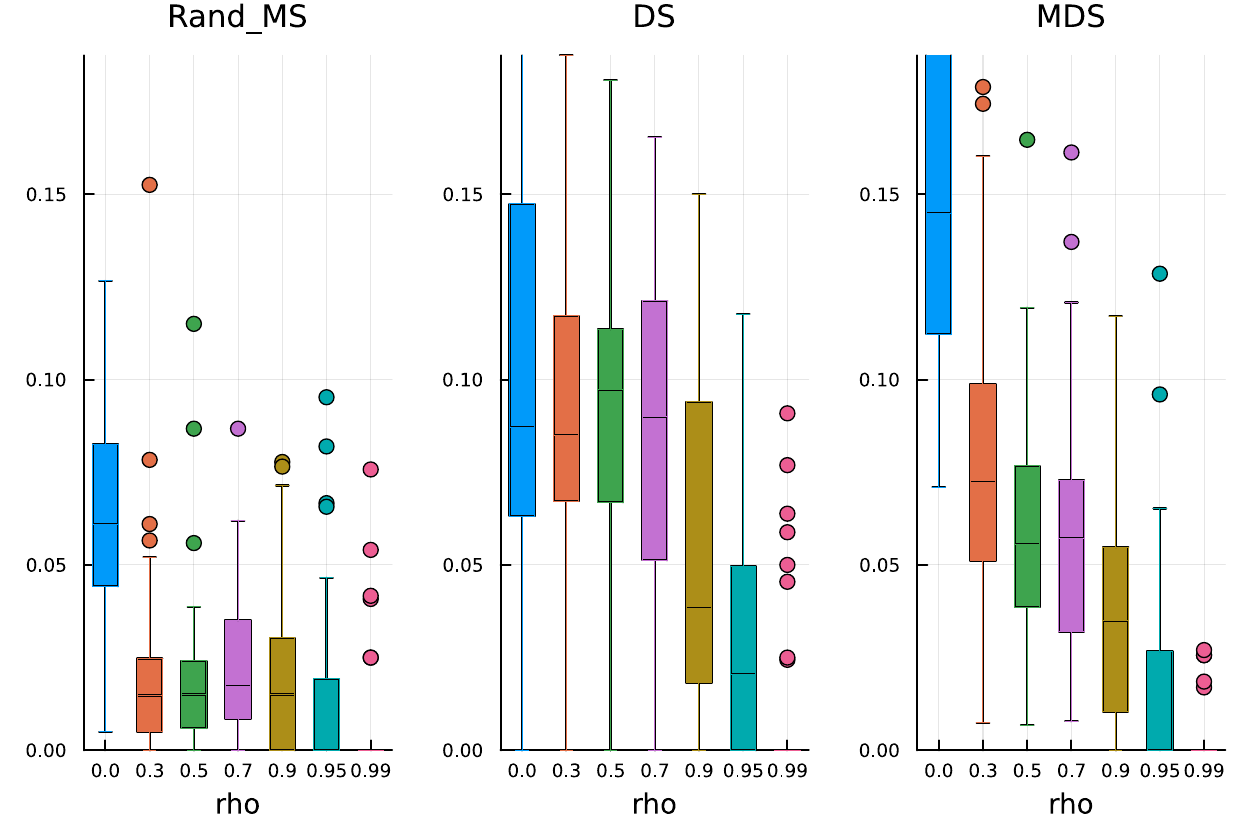}

    \vspace{0.5cm}
    
    \includegraphics[width=1\linewidth]{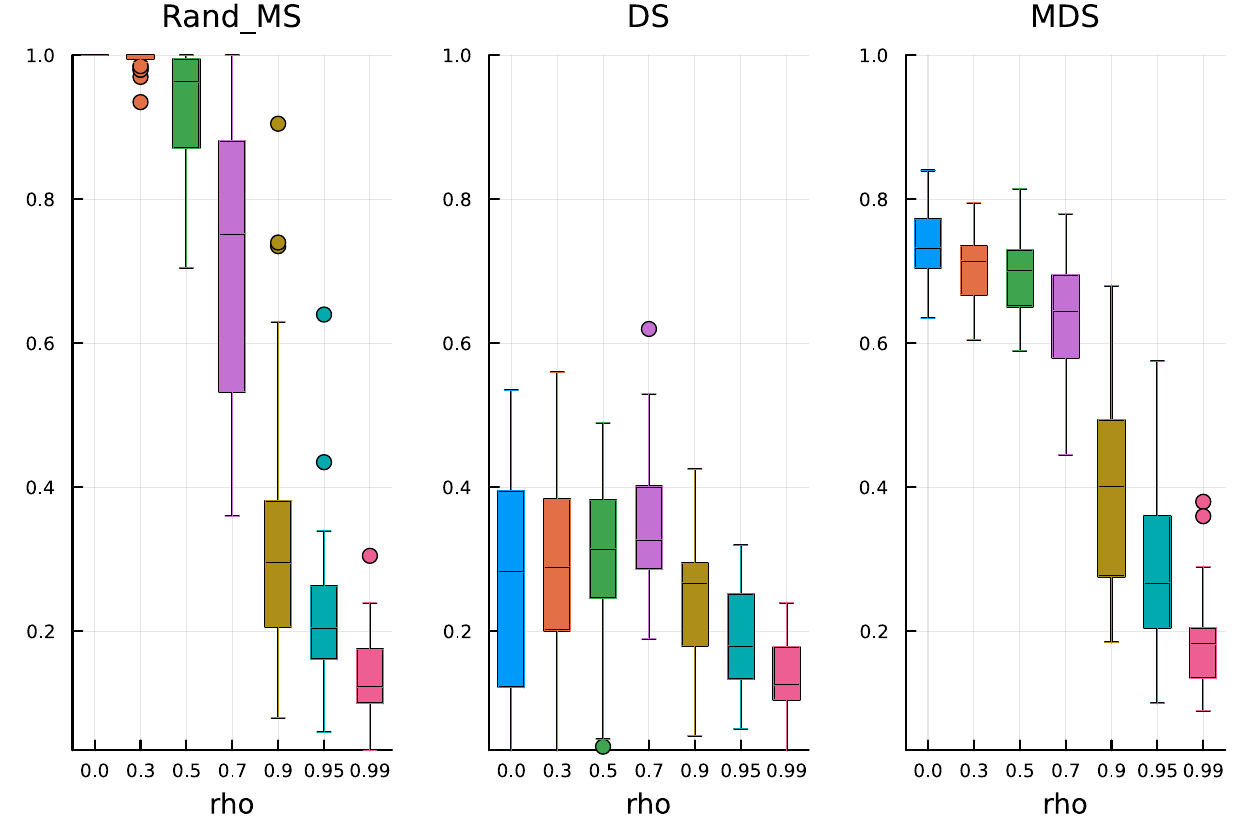}
    \caption{\textcolor{blue}{Active coefficients $20\%$}\\
        Top: False Discovery Rate\\
        Bottom: True Positive Rate}
    \label{fig:p20_lasso}
\end{figure}

By increasing the percentage of active variables to $20\%$, we can appreciate a sharper difference, at the advantage of RandMS, both in terms of FDR and TPR. In Figure \ref{fig:p20_lasso} (top) we see that DS and MDS start to loose control over the FDR, in particular MDS is no longer as conservative as before. RandMS is still able to achieve the required FDR control at the pre-specified level.\\
In Figure \ref{fig:p20_lasso} (bottom) we see that RandMS outperform the competitors in terms of TPR, in particular for correlations up to, and including, $0.5$. On the other hand, DS is totally unable to retain enough power.

\begin{figure}
    \centering
    \includegraphics[width=1\linewidth]{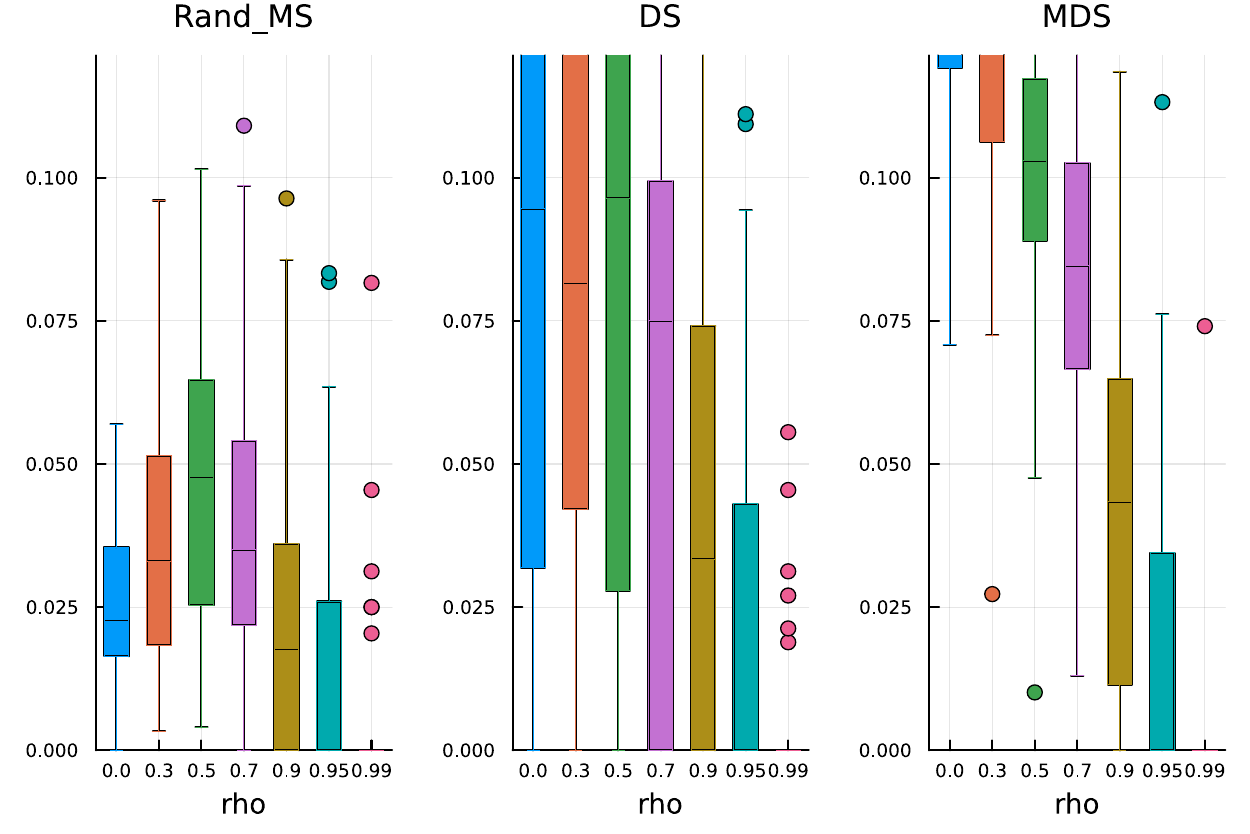}

    \vspace{0.5cm}
    
    \includegraphics[width=1\linewidth]{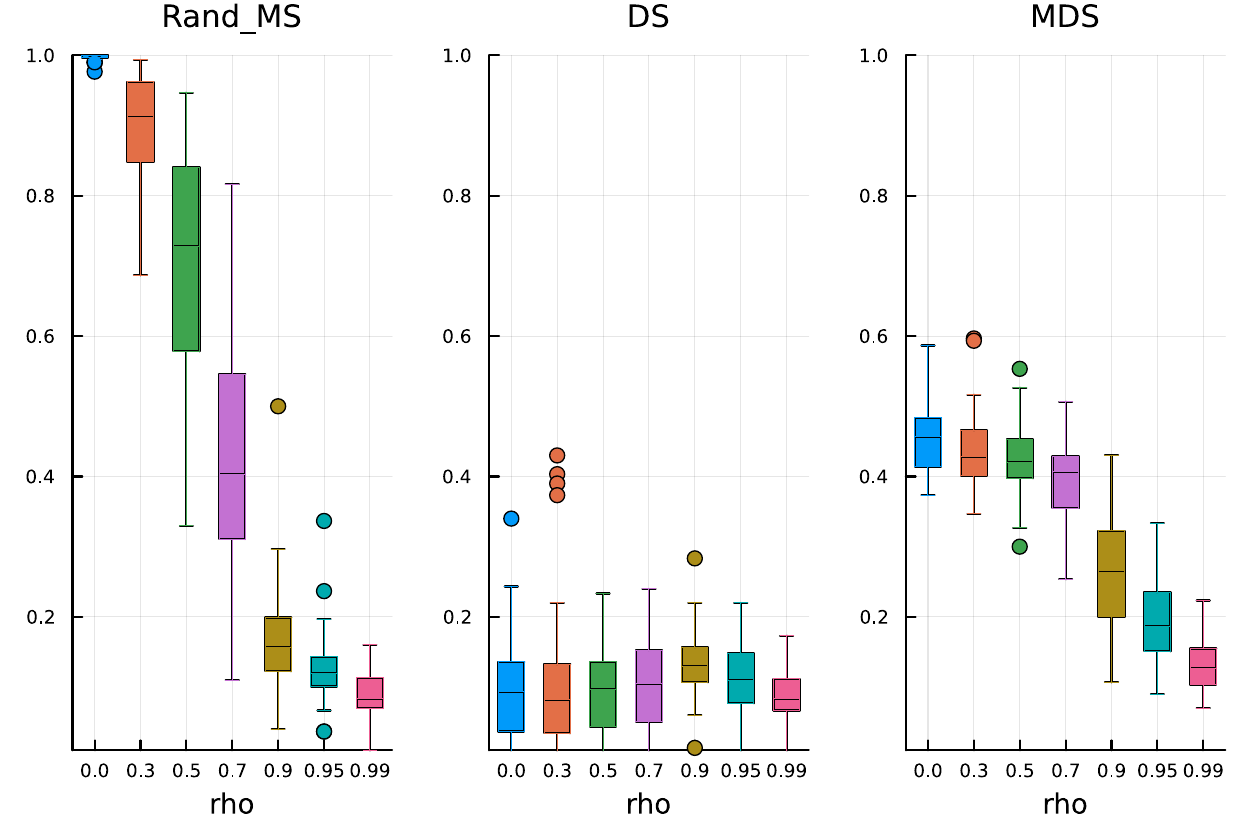}
    \caption{\textcolor{blue}{Active coefficients $30\%$}\\
        Top: False Discovery Rate\\
        Bottom: True Positive Rate}
    \label{fig:p30_lasso}
\end{figure}

Finally, setting the percentage of active variables to $30\%$, the difference is even more striking, again in favour of RandMS. Figures \ref{fig:p30_lasso} (top) and \ref{fig:p30_lasso} (bottom) show the results for FDR and TPR, respectively.

\subsubsection{Different covariance matrix structure}
\begin{figure}
    \centering
    \includegraphics[width=1\linewidth]{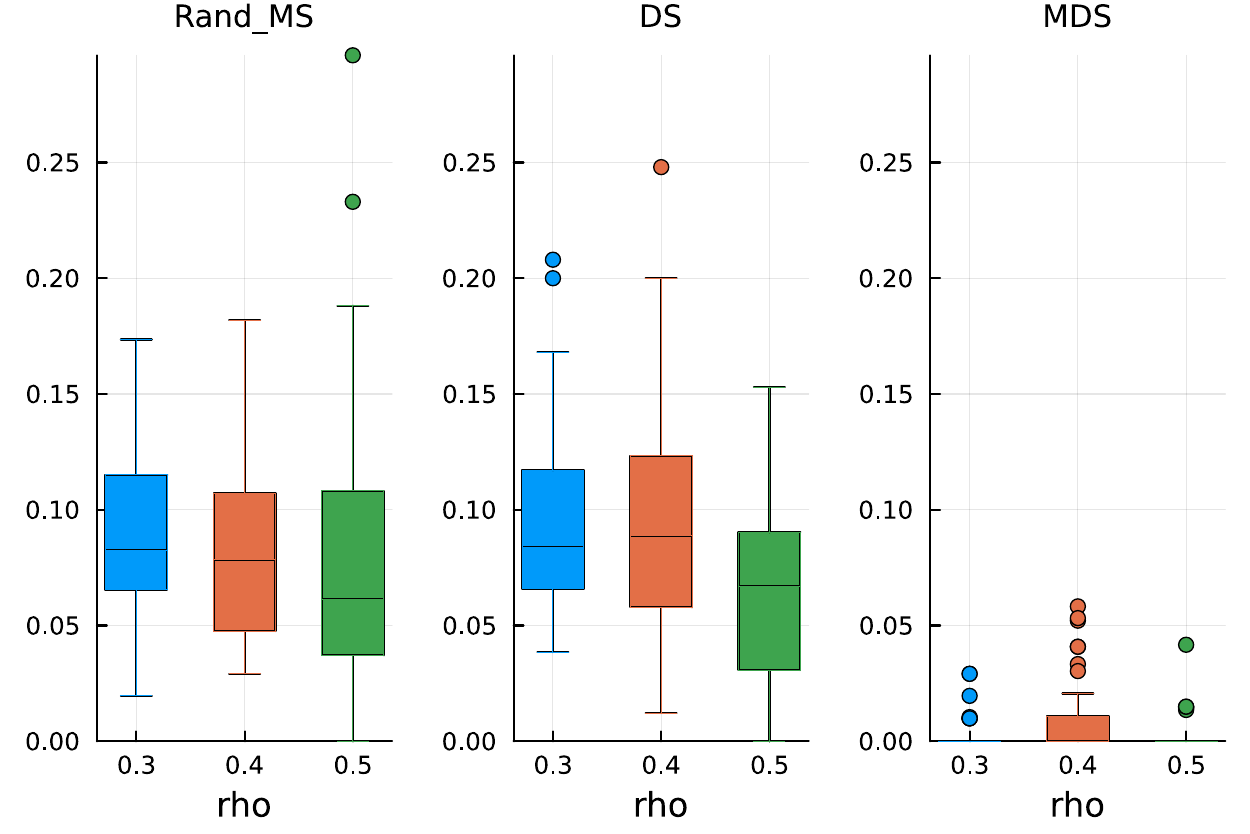}

    \vspace{0.5cm}
    
    \includegraphics[width=1\linewidth]{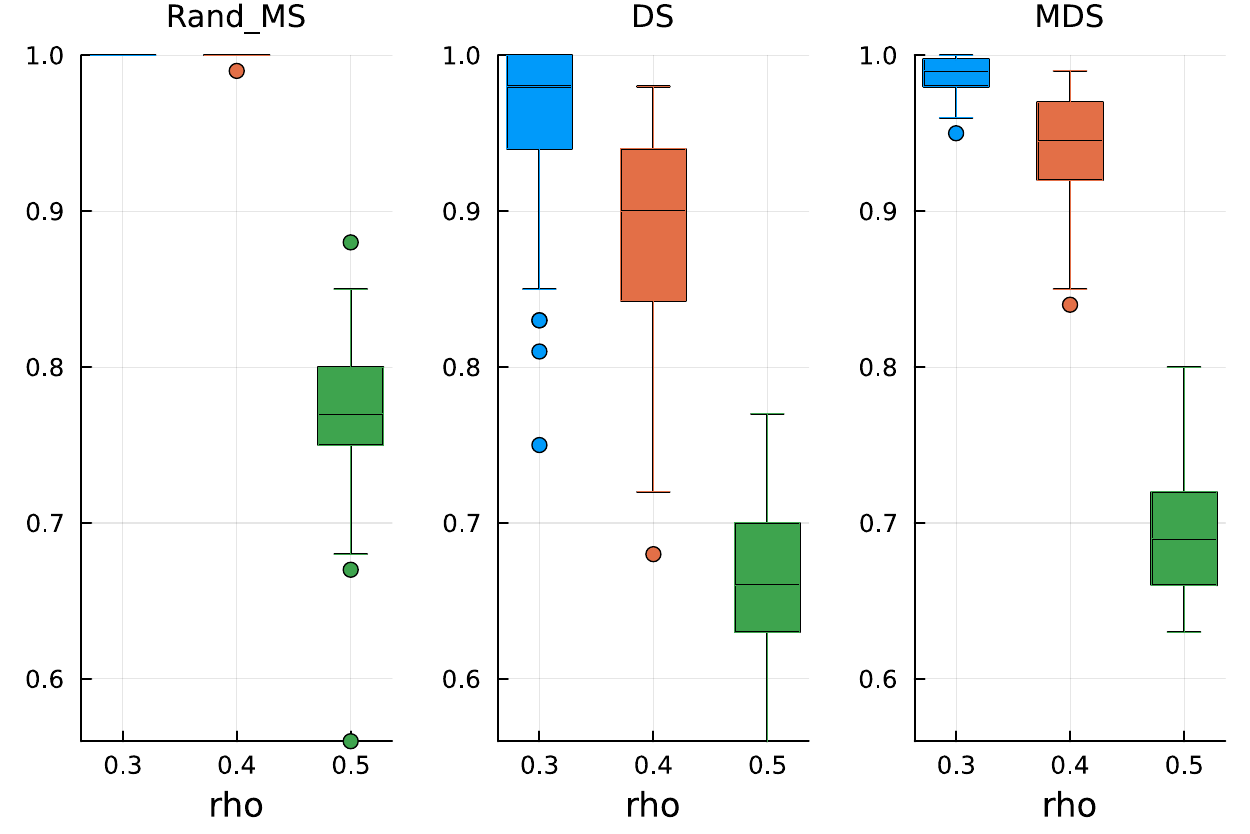}
    \caption{\textcolor{blue}{Ill conditioned covariance - active coefficients $10\%$}\\
        Top: False Discovery Rate\\
        Bottom: True Positive Rate}
    \label{fig:ill_cov_p10}
\end{figure}

\begin{figure}
    \centering
    \includegraphics[width=1\linewidth]{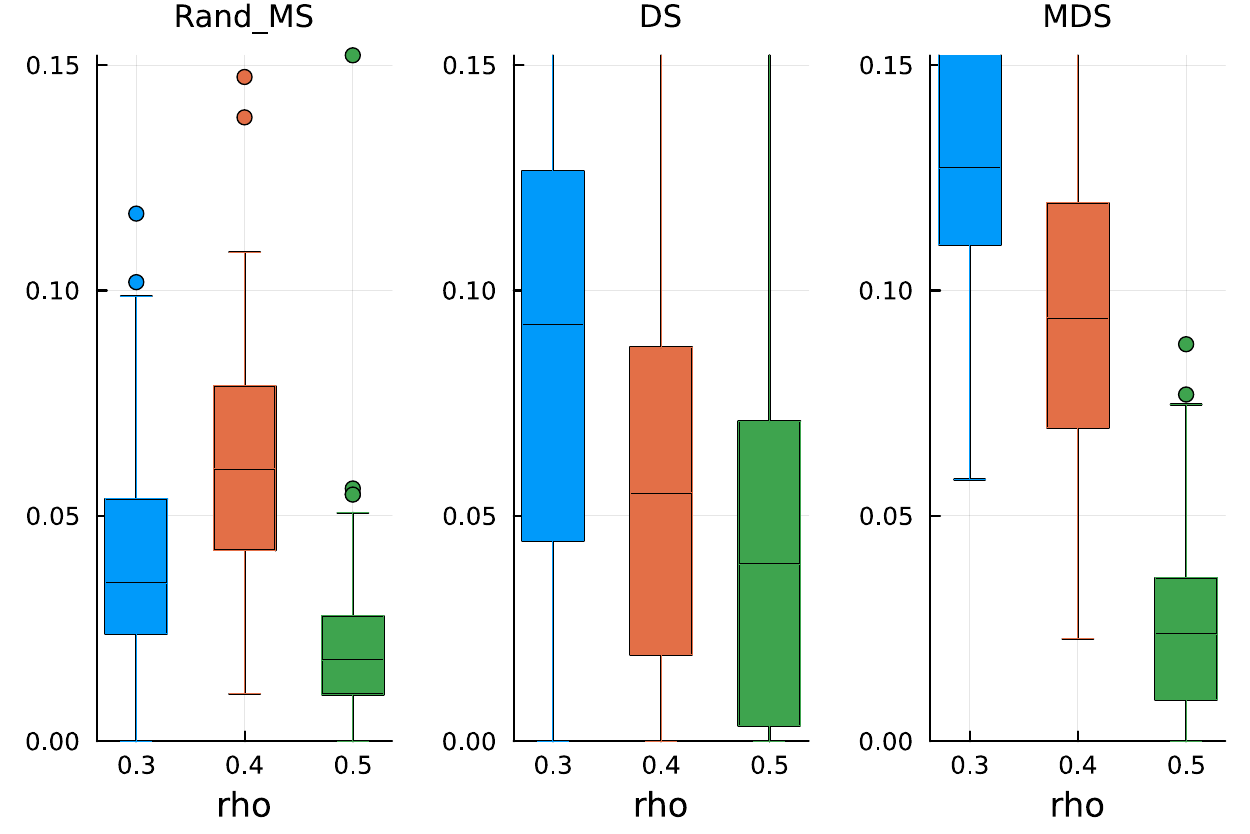}

    \vspace{0.5cm}
    
    \includegraphics[width=1\linewidth]{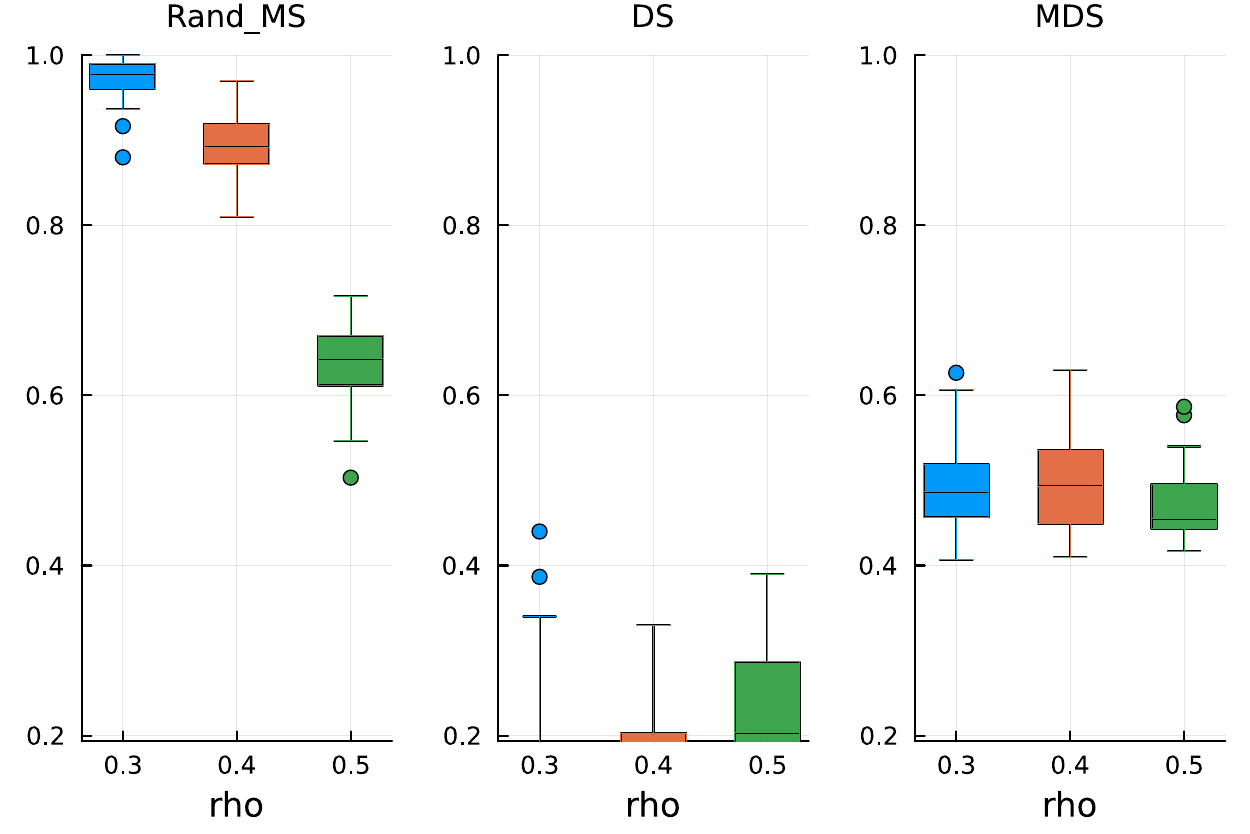}
    \caption{\textcolor{blue}{Ill conditioned covariance - active coefficients $30\%$}\\
        Top: False Discovery Rate\\
        Bottom: True Positive Rate}
    \label{fig:ill_cov_p30}
\end{figure}

The next simulation is performed with covariates generated from a Normal distribution whose inverse covariance matrix is near ill-conditioned, meaning that the lowest eigenvalues of $\Sigma^{-1}$ are nearly $0$. This results in a more unstable data generation.\\
The covariance matrix $\Sigma$ is constructed starting from the identity matrix and changing only the first off-diagonal entries to be equal to some specified values, here denoted by $\rho \in {0.3, 0.4, 0.5}$. This covariance structure implies that
\begin{align*}
    \text{corr}\left(X_j, X_{j+1}\right) &\ne 0\\
    \text{corr}\left(X_{j}, X_{j+2}\right) &= 0
\end{align*}
In Figure \ref{fig:ill_cov_p10} (top) we see that MDS is still conservative in terms of FDR, while DS and RandMS correctly control FDR at $10\%$. However, as we increase the percentage of active variables to $30\%$, Figure \ref{fig:ill_cov_p30} (top), we see that MDS is not conservative anymore, while RandMS still works well.\\
In terms of TPR, the advantage of RandMS is more clear. Already with a percentage of active coefficients of $10\%$, Figure \ref{fig:ill_cov_p10} (bottom), RandMS does a better job than DS and MDS, and, increasing that percentage to $30\%$, Figure \ref{fig:ill_cov_p30} (bottom), RandMS totally outperforms the competitors.

\subsubsection{Increasing $p$}
\begin{figure}
    \centering
    \includegraphics[width=1\linewidth]{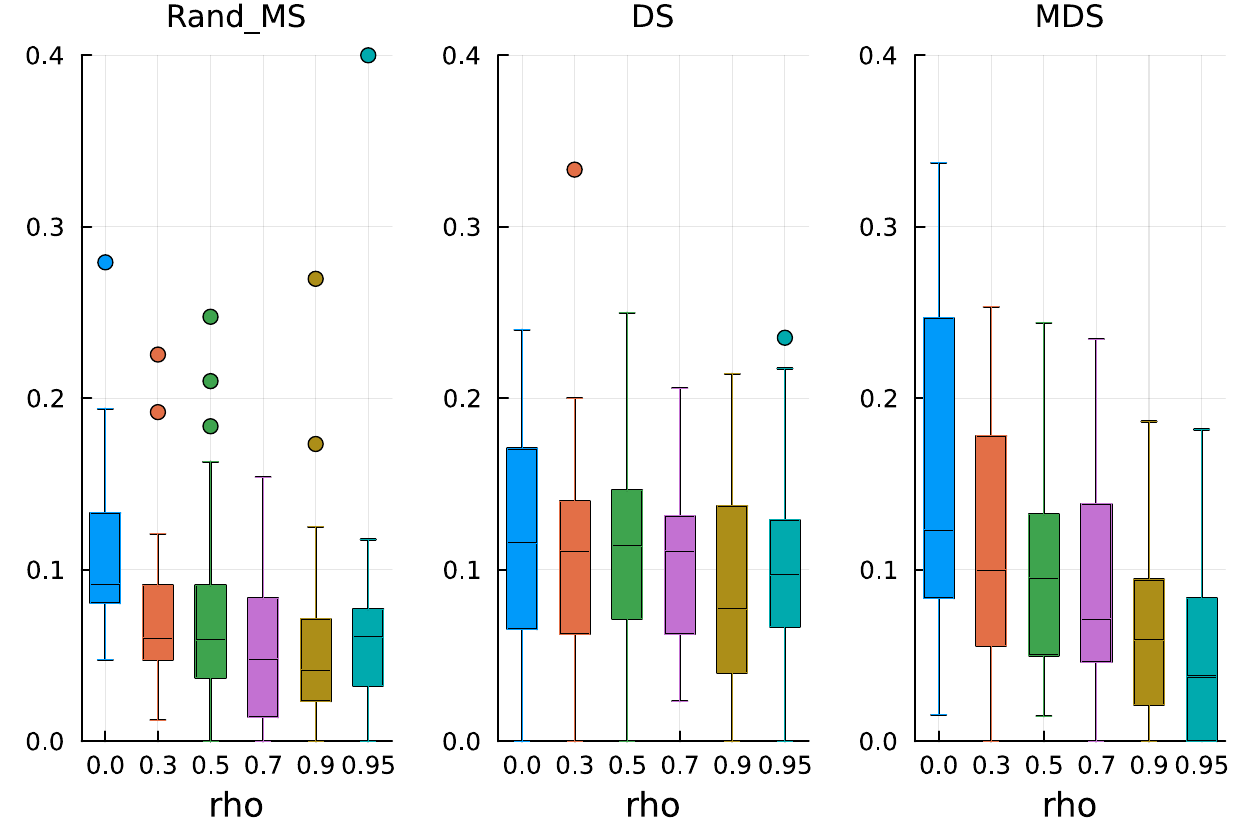}

    \vspace{0.5cm}
    
    \includegraphics[width=1\linewidth]{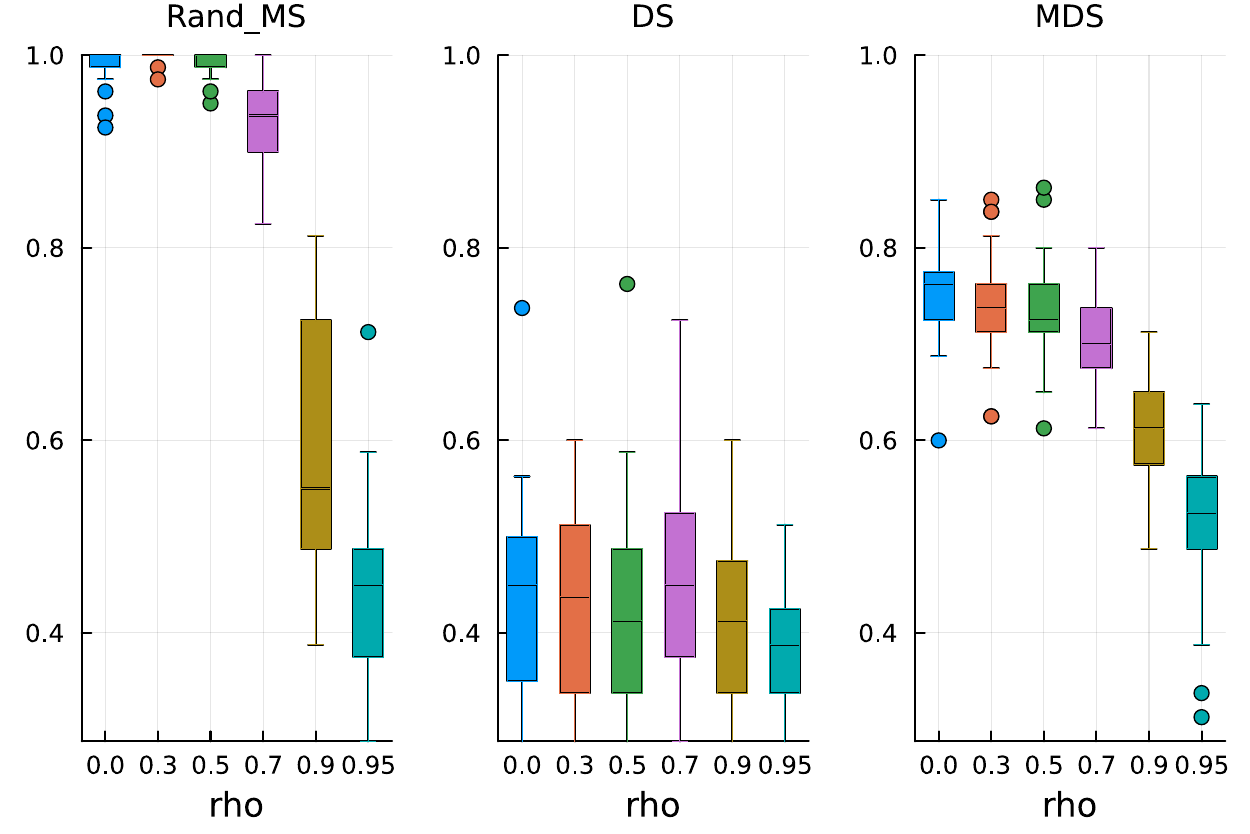}
    \caption{\textcolor{blue}{Block covariance matrix - $p = 10000$ with $p_1 = 80$}\\
        Top: False Discovery Rate\\
        Bottom: True Positive Rate}
    \label{fig:block_cov_p10000}
\end{figure}

Here we test the performance of RandMS with a higher number of covariates, $p=10000$, while keeping the sample size fixed at $n=800$. We repeat the simulations for different proportions of non-zero coefficients, from $0.003$ to $0.008$, which correspond to $p_1 = 30$ and $p_1 = 80$, respectively.\\
In Figure \ref{fig:block_cov_p10000} we report the performance for $p_1 = 80$. RandMS is able to control the FDR at $0.1$ as required and is comparatively better than the other methods that show a higher variability. The TPR is much higher for RandMS up to, and including, a correlation factor of $0.7$, with TPR values close to $1$. The performance decreases for more extreme correlation levels as expected.

\subsection{Computational performance}
\begin{figure}
    \centering
    \includegraphics[width=1\linewidth]{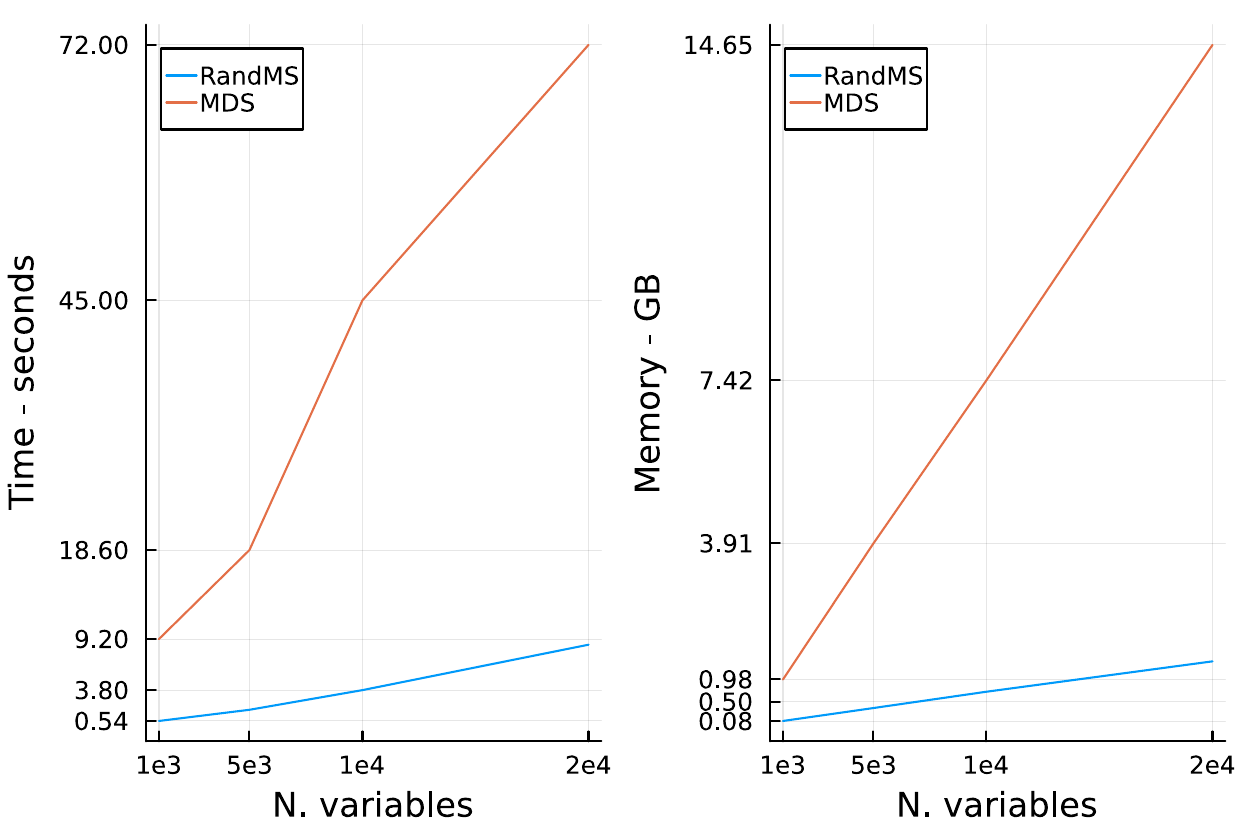}
    \caption{Computational benchmark of Randomisation with Mirror Statistic vs Multiple Data Splitting\\
    Left: CPU time in seconds\\
    Right: Memory requirement in Gigabytes}
    \label{fig:comp_time}
\end{figure}

We run a benchmark simulation of the computational requirements for the proposed Randomisation plus Mirror Statistic versus Multiple Data Splitting (run with $50$ splits). The benchmark, as well as all other analysis, has been done in \textit{Julia},\cite{Bezanson2017}, version $1.9.3$, on a Lenovo ThinkPad machine, with Linux OS and equipped with a \texttt{13th Gen Intel® Core™ i5-1345U × 12} CPU.\\
In Figure \ref{fig:comp_time} we show the average time (in seconds) and the memory requirements (in gigabytes) to run the algorithms on a linear regression with an increasing number of variables $p$, with a fixed number of active variables $p_1 = 30$, sample size $n=300$ and a block diagonal covariance matrix. For the Randomisation method both time and memory requirements scale linearly with the number of variables; as an example, with $p=10000$ the time required to estimate the full model is about $4$ seconds, while the memory required is about $0.7$ GB. In comparison, Multiple Data Splitting requires an order of magnitude more of time and memory. For example, with $p=10000$ variables, MDS takes on average $45$ seconds and requires over $7$ GBs of memory.

\section{Real data application to the identification of genes regulating fasting triglyceride levels}\label{sec:Application}
Elevated serum triglyceride (TG) levels in blood are strongly associated with increased risk of cardiovascular diseases (CVD). Serum TG levels can be reduced by a healthy diet, but there are also large inter-individual variation in fasting TG levels. An improved understanding of this variation would be beneficial for CVD prevention. In this example, we are interested in relating fasting TG levels to gene expression through a linear regression model. We have data from the screening visit of a randomized controlled dietary intervention trial, presented in detail in Ulven et al.\cite{Ulven2016}. In this trial, gene expression was measured in Peripheral blood mononuclear cells (PBMC). These are immune system cells and because they are circulating cells, they are exposed to nutrients, metabolites and peripheral tissues and may therefore reflect whole-body health. We include all individuals from whom we have both PBMC gene expressions and fasting TG levels, in total 251 individuals.
The outcome is the log measurement of blood triglycerides, while we use measurements from $13967$ genes expression as covariates ($HumanHT-12$ Expression BeadChips. Pre-processed gene expression probe level intensity values).\\
The log transformed triglyceride outcome is well approximated by a Normal distribution, while the gene expression data have been already preprocessed and are also well approximated by a Normal distribution.\\

We proceed to analyse the data with our proposed method, i.e. outcome Randomisation plus Mirror Statistic, in order to identify which genes could potentially contribute to the differences in TG levels. We use LASSO for variable selection on the randomised outcome $U$ and a standard linear regression model for the coefficients estimation on the randomised outcome $V$ (Algorithm \ref{alg:randms}); we set the randomisation parameter $\gamma = 1$ and we choose $f(a, b) = a + b$ for the test statistic in Equation \ref{eq:mirror_stat}. The FDR target level is set to $0.1$.\\
If we look only at the variable selection part of Algorithm \ref{alg:randms} (i.e. LASSO), the number of selected genes is $30$, while only $2$ of those (plus the intercept) have been selected using the full RandMS algorithm. Both our selected genes can be linked to atherosclerosis, and thus, seem biologically plausible. \textit{MYLIP} ($0.83 \left[0.74, 0.92\right]$) is related to TG level through LDL transport, while an over expression of \textit{ABCG1} ($0.89 \left[0.80, 0.98\right]$) results in increased efflux of cellular cholesterol to HDL, which has an inverse association with TG. The numbers in parentheses are the estimated multiplicative effect with the respective $95\%$ confidence interval in square brackets, thus an increased expression of both genes seem to have a positive effect on TG levels.

\section{Discussion}\label{sec:Discussion}
We propose the adoption of outcome randomisation instead of Data Splitting, in combination with the Mirror Statistic, in order to effectively control the False Discovery Rate in high-dimensional linear regressions. Intuitively, Randomisation acts as information averaging and helps avoid the pitfalls of Data Splitting. When combined with the Mirror Statistic, it allows to correctly control the FDR at the target level, while providing higher power and a more computationally efficient algorithm.\\
Our extensive simulations show the superior performance compared to Data Splitting strategies, in various scenarios of increasing complexity. Even in very high-dimensional cases we can retain good scalability of the proposed method.\\
Finally, we perform a real data analysis, where the outcome of interest is blood triglyceride levels, and the covariates are gene expression data. The dimension of the covariates space, compared to the sample size, makes this problem a perfect example of high-dimensional linear regression. We use our method to perform variable selection and inference, with a target FDR of $10\%$. We are able to identify two genes, potentially responsible for the variation of triglyceride levels.\\
This extension is currently limited to Normally distributed outcomes, where randomisation takes a closed form analytical solution and the symmetry requirement of the regression coefficients to apply the Mirror Statistic is satisfied. It is therefore possible to use the method for the inference on linear regression models and Gaussian graphical models, for example. It would be interesting to explore possible extensions to outcomes following arbitrary distributions and high-dimensional mixed models. Leiner et al.\cite{Leiner2023} provides an extension of randomisation to distributions belonging to the exponential family, relying on the concept of conjugate distributions from Bayesian statistics. Their result could potentially be used, for example for a high-dimensional logistic regression model. Dai et al.\cite{Dai2023} extend the use of the Mirror Statistic to high-dimensional logistic regression, however, their method relies on the computationally expensive procedure of de-biasing the LASSO estimate, needed to satisfy the symmetry requirement to use the Mirror Statistic. Future efforts to combine these two extensions could be of practical interest. Another area of further improvement could be the adoption of different variable selection techniques in Algorithm \ref{alg:randms}. For example, substituting LASSO with a model that can better select highly correlated covariates, e.g. ElasticNet\cite{Zou2005}.

\bibliographystyle{SageV}

\begin{acks}
The authors wish to thank Prof. Stine Ulven, Department of Nutrition, University of Oslo, for providing the Triglyceride dataset.
\end{acks}

\section*{Code availability}
The code used to perform all the simulations, the real data analysis and the computational performance benchmark is available at \url{https://github.com/marcoelba/SelectiveInference}.

\section{Appendix}\label{sec:appendix}
\begin{figure}[!ht]
    \centering
    \includegraphics[width=1\linewidth]{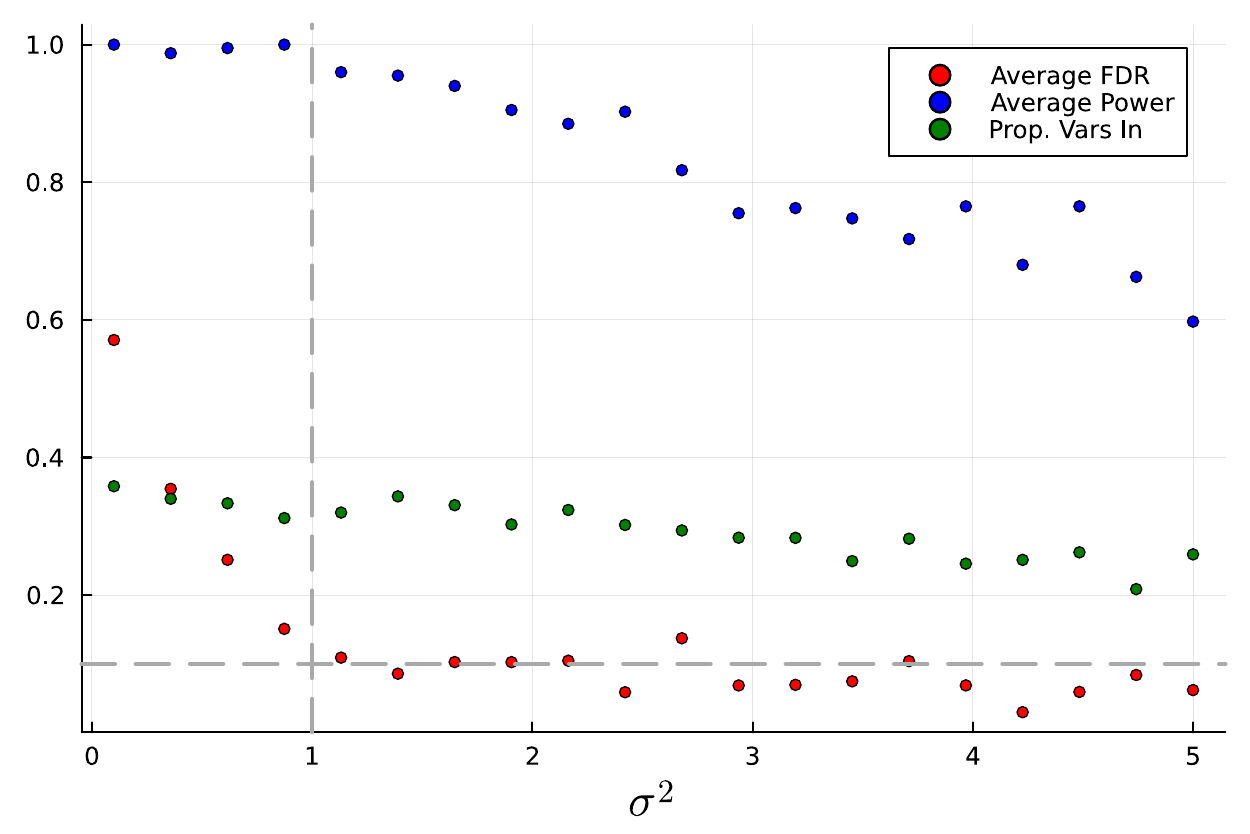}
    \caption{Average FDR, Power and proportion of selected variables against different values of ${\sigma}^2$}
    \label{fig:sigma_vs_fdr}
\end{figure}

We explore the implications of an incorrectly estimated variance ${\sigma}^2$ through additional simulations. We generate data for a linear regression model with ${\sigma}^2 = 1$ and we evaluate the performance of the algorithm for a range of values of ${\sigma}^2$, which are directly provided as input into the model, and a target FDR of $0.1$. In Figure \ref{fig:sigma_vs_fdr} we can see how different values of the variance affect our proposed strategy in terms of FDR and Power, as well as number of variables selected by the variable selection step (LASSO). For each value of ${\sigma}^2$ we average the metrics over $20$ generated datasets. When the variance is underestimated the FDR is not properly controlled. The estimated variance is needed for the randomisation of the outcome in Algorithm \ref{alg:randms} and if the variance is too small, relative to the ground-truth value, then $\boldsymbol{u}$ and $\boldsymbol{v}$ are not independent. As a consequence the variable selection becomes biased and overconfident.\\
When the variance is overestimated the algorithm becomes more conservative and the FDR is controlled. The number of variables selected stays roughly constant for the selected range of values. 

\begin{figure}[!ht]
    \centering
    \includegraphics[width=1\linewidth]{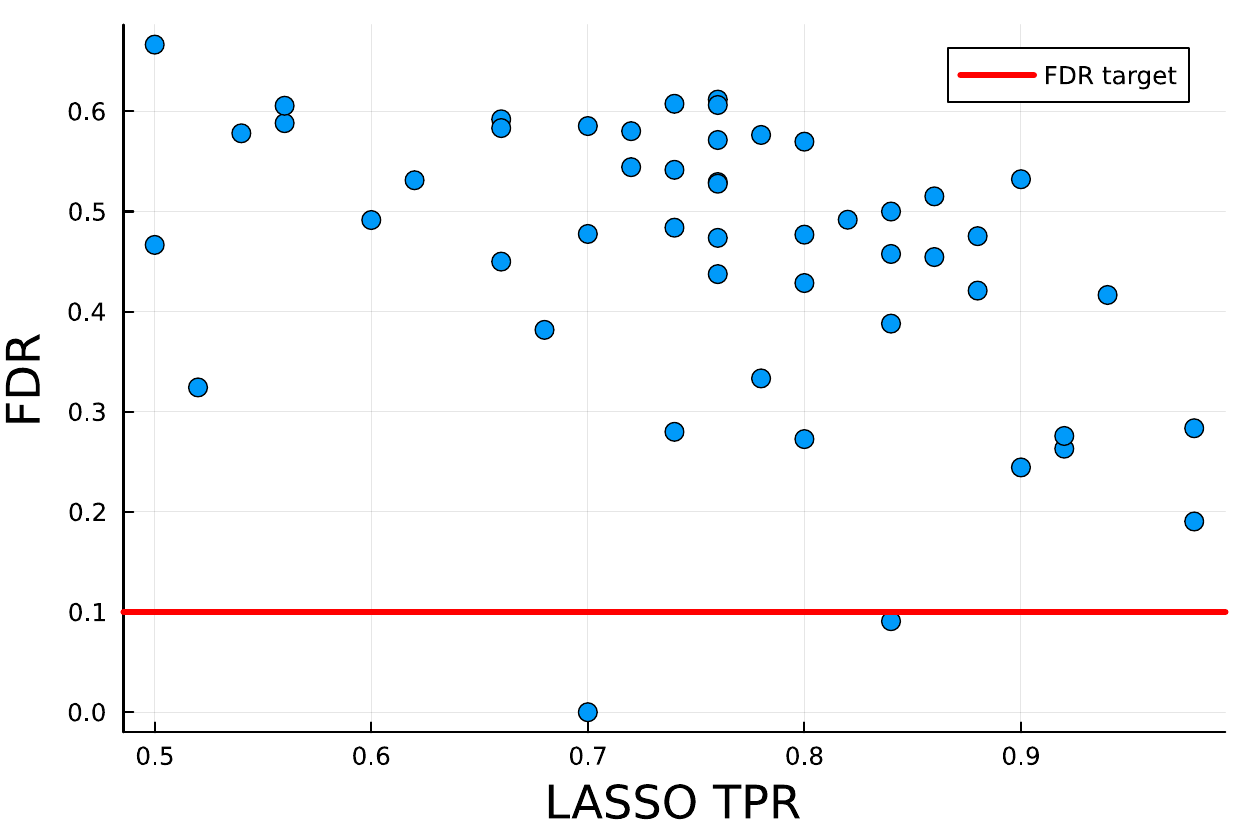}
    \caption{Proportion of true non-null variables selected by the LASSO (x-axis) against the FDR achieved with RandMS}
    \label{fig:lassotpr_vs_fdr_wrong}
\end{figure}

\begin{figure}[!ht]
    \centering
    \includegraphics[width=1\linewidth]{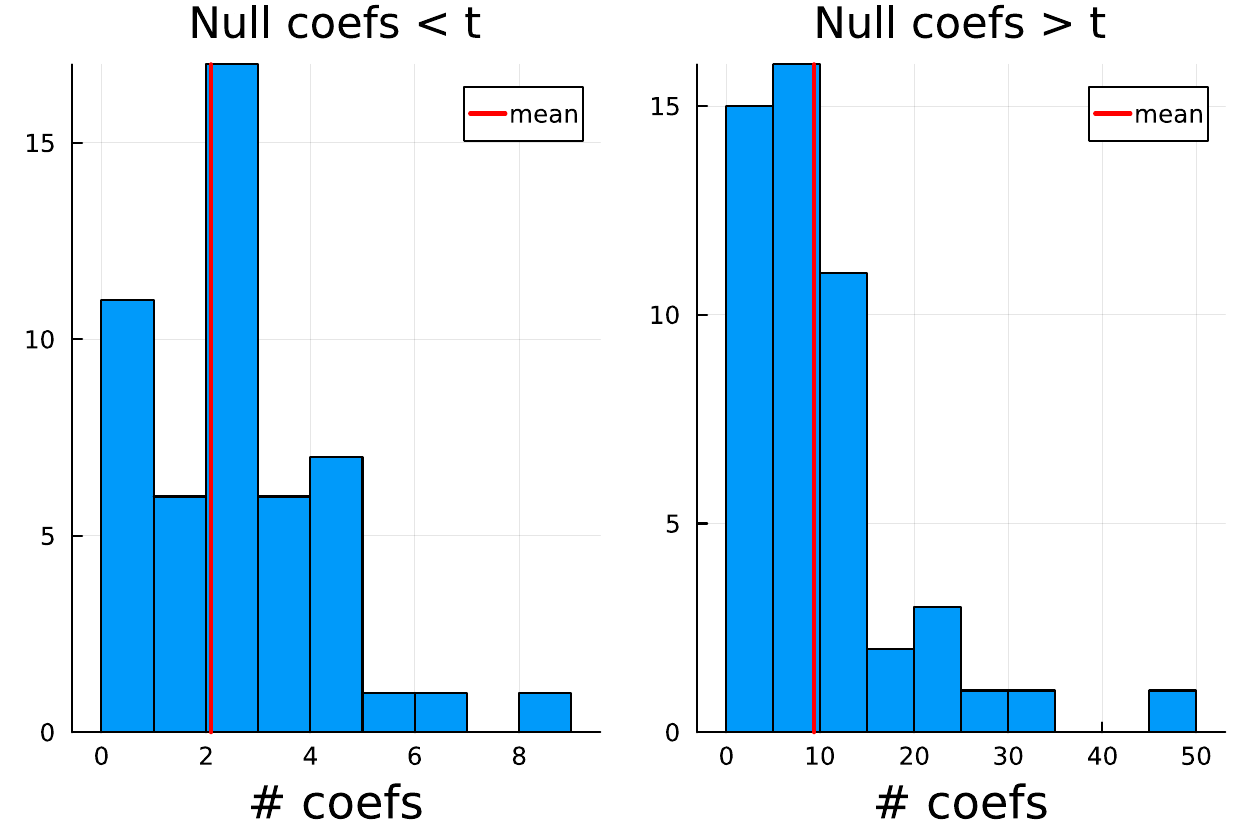}
    \caption{Distribution of the number of Mirror Statistic coefficients (associated to the true \textit{null}-coefficients) lying above and below the optimal threshold calculated for each repeated sampling.}
    \label{fig:NO_simmetry_histograms}
\end{figure}

In order to better understand the impact of violating the assumption of symmetry, we run a controlled simulation where we monitor the individual steps of the randomization plus Mirror Statistic algorithm. We generate data from a linear regression model with $n=200$ and $p=1000$ and $p_1 = 50$ non-null regression coefficients $\beta_j \in \{-1, 1\}$. The covariates are random draws from a multivariate Normal distribution with covariance matrix built according to Equation \ref{eq:sigma_toeplitz} with correlation factor $\rho=0.5$. The experiment is repeated $50$ times on repeated samples. By making the experiment hard on purpose we produce a LASSO step where not all active variables are selected, therefor nullifying the \textit{sure screening} property, which then cause the Mirror Statistic coefficients to not be distributed symmetrically around $0$. In Figure \ref{fig:lassotpr_vs_fdr_wrong} we compare the proportion of true non-null variables selected by the LASSO (x-axis) against the FDR achieved with RandMS. It is clear how the violation of the \textit{sure screening} property affects the final FDR. In Figure \ref{fig:NO_simmetry_histograms} we actually see how this is caused by a violation of the symmetry requirement, which is fundamental for the approximation in Equation \ref{eq:fp_approximation} to work. If the symmetry assumption were fulfilled, then the number of coefficients (associated to the true \textit{null}-regression coefficients) above and below the optimal threshold calculated through Algorithm \ref{alg:randms} should be roughly the same, while this is clearly not the case for this experiment.

\end{document}